\documentclass[twocolumn]{aastex63}

\usepackage{hyperref}

\accepted{September 5, 2023}
\submitjournal{AAS}

\shorttitle{The 50 Mpc Galaxy Catalog}
\shortauthors{Ohlson et al.}

\graphicspath{{./}{figures/}}

\begin{document}

\title{The 50 Mpc Galaxy Catalog (50MGC): Consistent and Homogeneous Masses, Distances, Colors, and Morphologies}

\correspondingauthor{David Ohlson, david.ohlson@utah.edu}

\author[0009-0004-9457-2495]{David Ohlson}
\affiliation{University of Utah \\
201 James Fletcher Building\\
115 S. 1400 E.\\
Salt Lake City, UT 84112, USA}

\author[0000-0003-0248-5470]{Anil C. Seth}
\affiliation{University of Utah \\
201 James Fletcher Building\\
115 S. 1400 E.\\
Salt Lake City, UT 84112, USA}

\author[0000-0001-5802-6041]{Elena Gallo}
\affiliation{University of Michigan}

\author[0000-0003-4703-7276]{Vivienne F. Baldassare}
\affiliation{Washington State University}

\author[0000-0002-5612-3427]{Jenny E. Greene}
\affiliation{Princeton University}

\begin{abstract}
We assemble a catalog of 15424 nearby galaxies within 50 Mpc with consistent and homogenized mass, distance, and morphological type measurements.  Our catalog combines galaxies from HyperLeda, the NASA-Sloan Atlas, and the Catalog of Local Volume Galaxies.  Distances for the galaxies combine best-estimates for flow-corrected redshift-based distances with redshift independent distances. We also compile magnitude and color information for 11740 galaxies.  We use the galaxy colors to estimate masses by creating self-consistent color -- mass-to-light ratio relations in four bands; we also provide color transformations of all colors into Sloan $g-i$ by using galaxies with overlapping color information.   We compile morphology information for 13744 galaxies, and use galaxy color information to separate early and late-type galaxies.  
This catalog is widely applicable for studies of nearby galaxies, and placing these studies in the context of more distant galaxies.  We present one application here; a preliminary analysis of the nuclear X-ray activity of galaxies. Out of 1506 galaxies within the sample that have available Chandra X-ray observations, we find 291 have detected nuclear sources. Of the 291 existing Chandra detections, 249 have log(L$_{X}$)$>$38.3 and available stellar mass estimates. We find that the X-ray active fractions in early-type galaxies are higher than in late-type galaxies, especially for galaxy stellar masses between 10$^9$ and 10$^{10.5}$~M$_\odot$.  We show that these differences may be due at least in part to the increased astrometric uncertainties in late-type galaxies relative to early-types.  
\end{abstract}


\section{Introduction} \label{sec:intro}

The galaxies in the nearby Universe provide our best laboratory for studying a wide range of scientific questions.  Most fundamentally, the properties of nearby galaxies are the outcome of the process of galaxy evolution over the lifetime of our Universe.  We can use our detailed view of nearby galaxies to quantify their morphology, kinematics, and star formation histories.

Our most complete census of galaxies comes from the local measurements.  Only within the Local Group can the faintest galaxies be found \citep[e.g.][]{willman05,mcconnachie12,simon19}, and even with modern surveys like the Sloan Digital Sky Survey (SDSS), we can get a census of bright dwarf galaxies with masses ($M_\star$) $\sim$10$^9$~M$_\odot$ only out to $\sim$50~Mpc \citep[e.g.][]{blanton05}.  The detailed structures within nearby galaxies can reveal many aspects of their history and evolution,  including their faint stellar halos \citep[e.g.][]{crnojevic16}, star cluster populations \citep[e.g.][]{krumholz19,hughes21}, dwarf galaxy satellites \citep[e.g.][]{carlsten22}, central stellar structures \citep[e.g.][]{erwin15,neumayer20}, and their supermassive black holes \citep[e.g.][]{saglia16}.
Nearby galaxies also provide a unique laboratory for stellar and high-energy astrophysics, such as quantifying rare stages of stellar evolution that aren't represented in the Milky Way \citep[e.g.][]{melbourne12}, finding and characterizing rare sources like ultraluminous X-ray sources \citep{kaaret17}, and transient events including gravitational wave events \citep{abbott17} and rare types of supernovae \citep[e.g.][]{tartaglia18}.  Finally, nearby galaxies provide a baseline for distance measurements in the Universe and measurement of their distances are key for understanding cosmology \citep[e.g.][]{freedman21}.  

A number of existing catalogs of nearby galaxies are regularly used in selecting sources for these science explorations.  The \citet{karachentsev13} Local Volume Galaxy catalog\footnote{\href{https://www.sao.ru/lv/lvgdb/introduction.php}{\tt https://www.sao.ru/lv/lvgdb/introduction.php}} is a regularly updated and very complete catalog of the nearest galaxies with distance $<$11~Mpc and includes distance and photometric information.  Beyond this distance, the most widely used sources of galaxy information are HyperLeda\footnote{\href{http://leda.univ-lyon1.fr/}{\tt http://leda.univ-lyon1.fr/}} \citep{makarov14} and the NASA Extragalactic Database (NED)\footnote{\href{http://ned.ipac.caltech.edu/}{\tt http://ned.ipac.caltech.edu/}}. In addition, many studies make use of catalogs of the nearest rich clusters of galaxies, Virgo \citep{binggeli85,kim14}.  Distance measurements to nearby galaxies has been a subject of intense study, and much of this information for nearby galaxies has been compiled within a subsection of NED \citep{steer17}.  All of these works combine heterogeneous photometric, distance, and morphological measurements, and make some attempt to synthesize and homogenize these.  Modern studies of galaxies typically characterize galaxies based on their stellar masses and colors \citep[e.g.][]{baldry04}.  Unfortunately, there is no uniform and homogeneous estimates of these quantities for nearby galaxies, and the varying photometric data as well as heterogeneous distance estimates available for these galaxies complicates their comparison to more distant galaxy samples.  

The goal of this paper is to provide the first homogenized source of nearby galaxy masses, distances, colors, and morphologies.  We have chosen a distance limit of $\sim$50~Mpc for this catalog, as many of the existing studies of nearby galaxies are limited to galaxies within this distance limit. We find in \S\ref{sec:mass} that these galaxies have stellar masses that range from log($M_{\star}/M_\odot$) of 6.6 to 10.9 (at the 99\%ile). We also present one application that motivated the creation of this catalog; a comparison of nuclear X-ray activity between early and late-type galaxies.  These results are the first step to determining constraints on the occupation fraction of central massive black holes in nearby galaxies \citep{miller15,gallo19}, which will be presented in a follow-up paper (Gallo et al., {\em in prep}). 

This paper is organized as follows: we describe our sample selection and photometry in \S\ref{sec:sample}.  We discuss distance estimates to the sample galaxies in \S\ref{sec:distance}, determine their masses based on consistent color-mass-to-light ratio measurements in \S\ref{sec:mass}, and describe their morphologies and group membership in \S\ref{sec:type}.  Finally in \S\ref{sec:xray}, we apply the catalog by combining it with Chandra archival data to present our X-ray active fraction measurement before concluding.  The full galaxy catalog is in the Appendix in Table~\ref{tab:fullcatalog}.  This table is available through the publisher and (potentially updated) at \href{https://github.com/davidohlson/50MGC}{\tt https://github.com/davidohlson/50MGC}.  Throughout the paper, when we refer to column names without other clarifying information, we refer to the full catalog; columns are indicated through their font, i.e.~{\tt column\_font\_format}.

\section{Sample Selection \& Photometry} \label{sec:sample}

In this section, we detail the construction of our sample from multiple sources as well as the compilation of photometry that enables luminosity and mass estimates for each galaxy.

\subsection{Base Catalogs \& Combination} \label{subsec:sample sources}
Our catalog is formed from three base sources, the Local Volume Galaxy (LVG) catalog \citep{karachentsev13}, HyperLeda \citep{makarov14}, and the NASA-Sloan Atlas (NSA) \footnote{\href{https://www.sdss.org/dr13/manga/manga-target-selection/nsa/}{\tt https://www.sdss.org/dr13/manga/manga-target-selection/\\nsa/}}.  All three contain a wide variety of information including redshifts and photometry.  The strength of the NSA catalog is its uniform Sloan Digital Sky Survey photometry extending to quite faint sources, however, this photometry only extends over $\sim$1/3rd of the sky.   The LVG and HyperLeda catalogs combine hetereogenous data sets and are continuously maintained and updated.  The LVG catalog contains nearly complete measurements of galaxies brighter than $M_B \sim -12$ within 10~Mpc, however it lacks optical color information for the galaxies.  Both NSA and HyperLeda extend to greater distances and have optical color information.  Combined, these catalogs can provide the data needed to obtain distances, masses, and Hubble types for galaxies out to maximum target distance of 50~Mpc.

From HyperLeda, we initially began with a subsample of 18726 objects designated as galaxies and with radial velocity $< 3500$~km~s$^{-1}$. We chose to remove the $\sim$20\% of objects which were missing absolute B-band magnitude values in column {\tt mabs} from our sample, because HyperLeda's calculation of absolute magnitude relied upon apparent B-band magnitude, distance modulus, and extinction. Removal of these objects from the sample ensured the inclusion of the most essential galaxy measurements. This cut the initial HyperLeda sample down to 14941 objects.

For the LVG sub-sample, we combined the "Catalog of Nearby Galaxies" and "Global Parameters of the Nearby Galaxies" tables available from the LVG catalog website\footnote{\href{http://www.sao.ru/lv/lvgdb/tables.php}{\tt http://www.sao.ru/lv/lvgdb/tables.php}}. The combined tables totalled 1240 galaxies\footnote{The version used was accessed 8/4/2020}.

Finally for the NSA data, we used the {\tt nsa\_v1\_0\_1} catalog and found 8242 objects with $cz<3500$~km~s$^{-1}$. We also added available SDSS spectral line flux data to these objects, matched by plate and fiber IDs.  We note that the NSA catalog includes not just objects with SDSS redshift information, but also galaxies with redshifts measured from other sources (ALFALFA, NED, 6dF, 2dF, and ZCAT).

\subsubsection{Combining the Sample} \label{subsec:combining}

The three subsamples contain both duplicate objects and contaminants. Here, we describe how we combined the catalogs while eliminating duplicates and contaminants.  We note that the presence of a galaxy in each input catalog is given by flags in our catalog: {\tt hl\_obj}, {\tt lvg\_obj}, and {\tt nsa\_obj}.  If a galaxy is present in that catalog the flag has a value of 1, while if it is not, it has a 0 value.

The HyperLeda sample included some duplicate galaxy entries; to remove these we crossmatched each entry to its nearest neighbor within the sample. We followed up on galaxies with nearby neighbors within a 2\arcmin~ separation radius for galaxies within 20 Mpc, and a 1\arcmin~ separation for galaxies between 20-50 Mpc.  For the 580 potential duplicate galaxies we examined available data and flagged them as follows --  0: individual galaxy with the best available data, 1: duplicated galaxy with less available data, 2: off-nuclear position in duplicate galaxy, 3: image shows a bright or star-forming region within a larger galaxy. We include only objects flagged with 0, which removed 183/580 galaxies. This cut resulted in an initial HyperLeda sample of 14758 galaxies.

We first created a combined LVG and HyperLeda sample. We used {\tt astropy.coordinates.SkyCoord} to conduct a coordinate-based crossmatch with a maximum separation distance of 30\arcsec. Of the 1240 galaxies in the LVG, 744 were found in HyperLeda.  The 496 unmatched LVG galaxies are almost entirely very low-luminosity. This explains their exclusion, as these galaxies are likely to lack the photometric and redshift measurements required to have been included in our initial HyperLeda sample. When combining sources between these catalogs, we prioritize the LVG galaxy properties in almost all cases, i.e.~radial velocities, $B$ band magnitudes, and distances (see \S\ref{sec:distance}).

We then crossmatched NSA catalog objects to our combined HyperLeda and LVG samples using a maximum separation distance of 30\arcsec. Of the 8242 NSA objects, 6966 matched the combined HyperLeda \& LVG catalog.  For the  the 1276 unmatched objects, we found that many were contaminants. To distinguish these contaminants from real galaxies, we visually inspected each galaxies' SDSS imaging and compared redshifts to those listed in the NASA Extragalactic Database (NED). At the lowest redshifts, the contaminants included foreground stars on top of image artifacts or distant galaxies. At larger distances, bright portions of galaxies with separate spectroscopic measurements from SDSS were often included as separate catalog entries. Some contaminants also were just image artifacts, and a small number contained no obvious image source whatsoever.  We flagged each of the 1276 unmatched NSA objects, based on our findings -- 1: image shows a bright or star-forming region within a larger galaxy. 2: image shows no object, an imaging artifact, a star, or a foreground star over a high redshift galaxy. 3: Duplicates of otherwise good galaxies, of which we flagged the entry with the most reliable redshift source to keep, in order: SDSS, NED, ZCAT. 4: Good imaging of a galaxy with redshift matching NED.  These qualtity flags are given for all 1276 galaxies in Table~\ref{tab:nsaflags}. Only objects in category 4 are considered as possible unique new NSA galaxies to add to the catalog.

Finally, to minimize the potential for duplicate galaxies, we wanted to ensure none of the unique NSA objects in category 4 were observations of the edge of a galaxy already included in our sample. To do this, we first needed uniform angular diameter measurements for each sample galaxy.

\begin{figure}
    \plotone{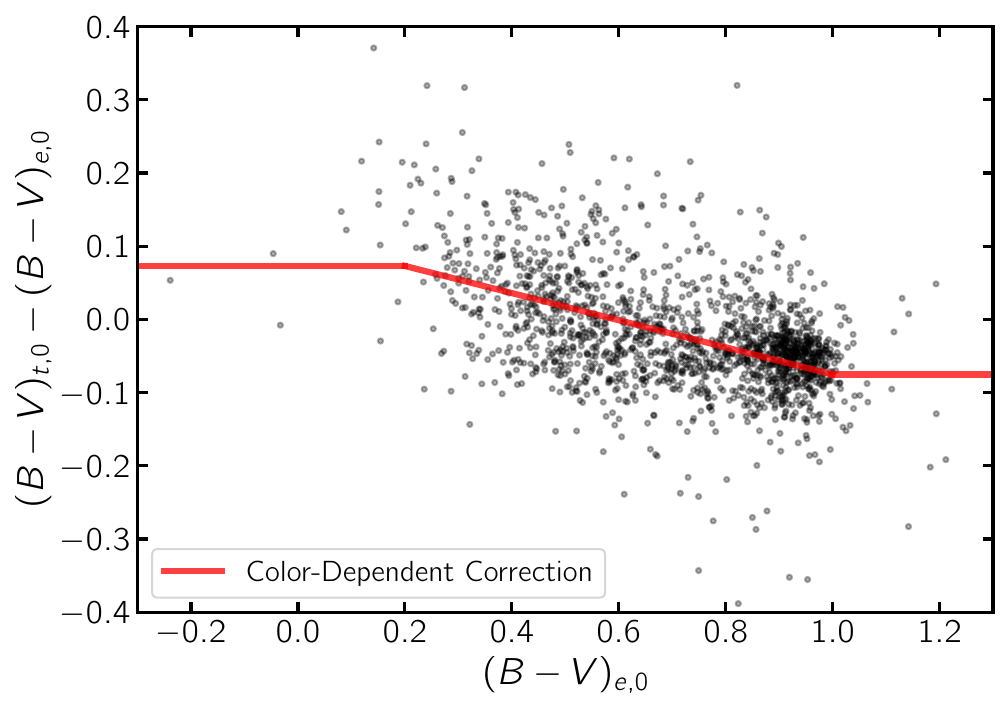}
    \caption{Difference between HyperLeda galaxies with overlapping total ($(B-V)_{t,0}$) and effective ($(B-V)_{e,0}$) colors vs their $(B-V)_{e,0}$ value. The red line shows a linear correction applied to $(B-V)_{e,0}$ values to correct for the slight color-dependent offset to allow combination of colors into a single field. Correction coefficients are based on a linear fit from $0.2\leq(B-V)_{e,0}\leq1.0$.
    \label{fig:bvcolor}}
\end{figure}

\begin{figure*}
\gridline{\fig{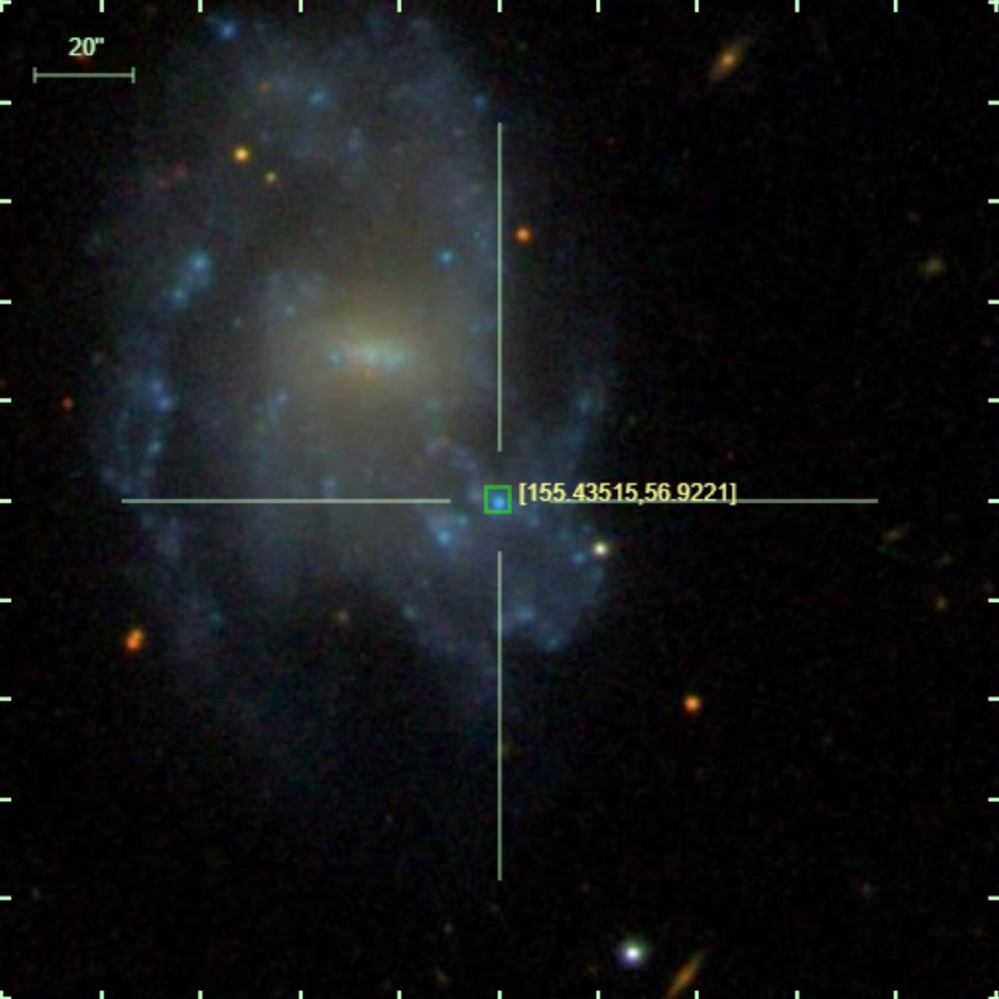}{0.24\textwidth}{(a)}
          \fig{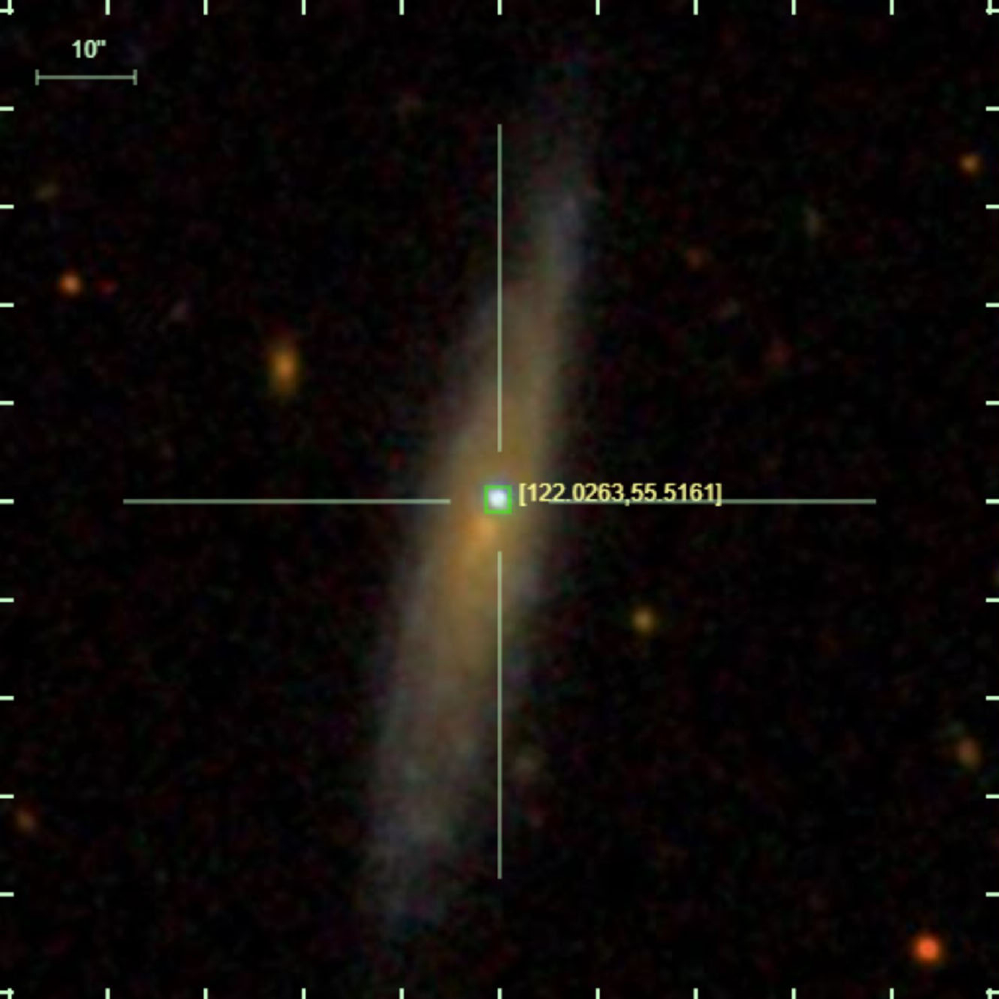}{0.24\textwidth}{(b)}          \fig{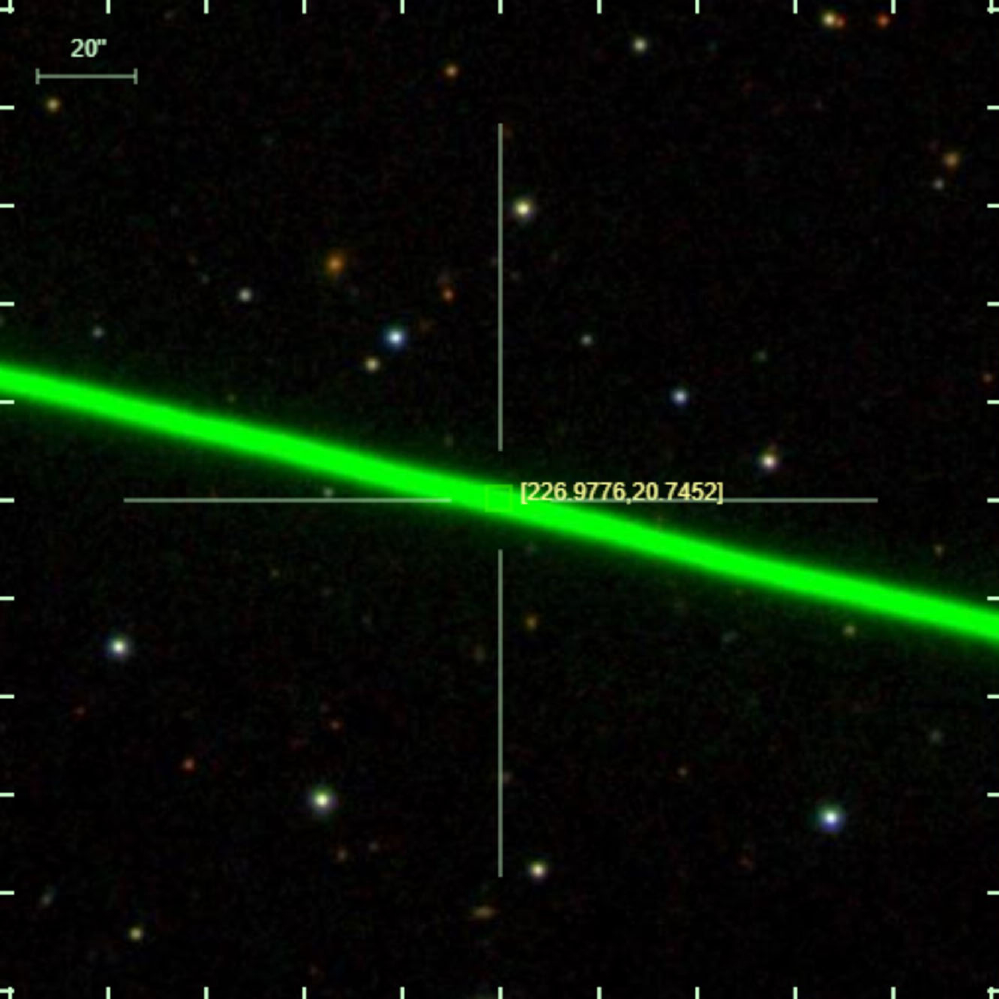}{0.24\textwidth}{(c)}
          \fig{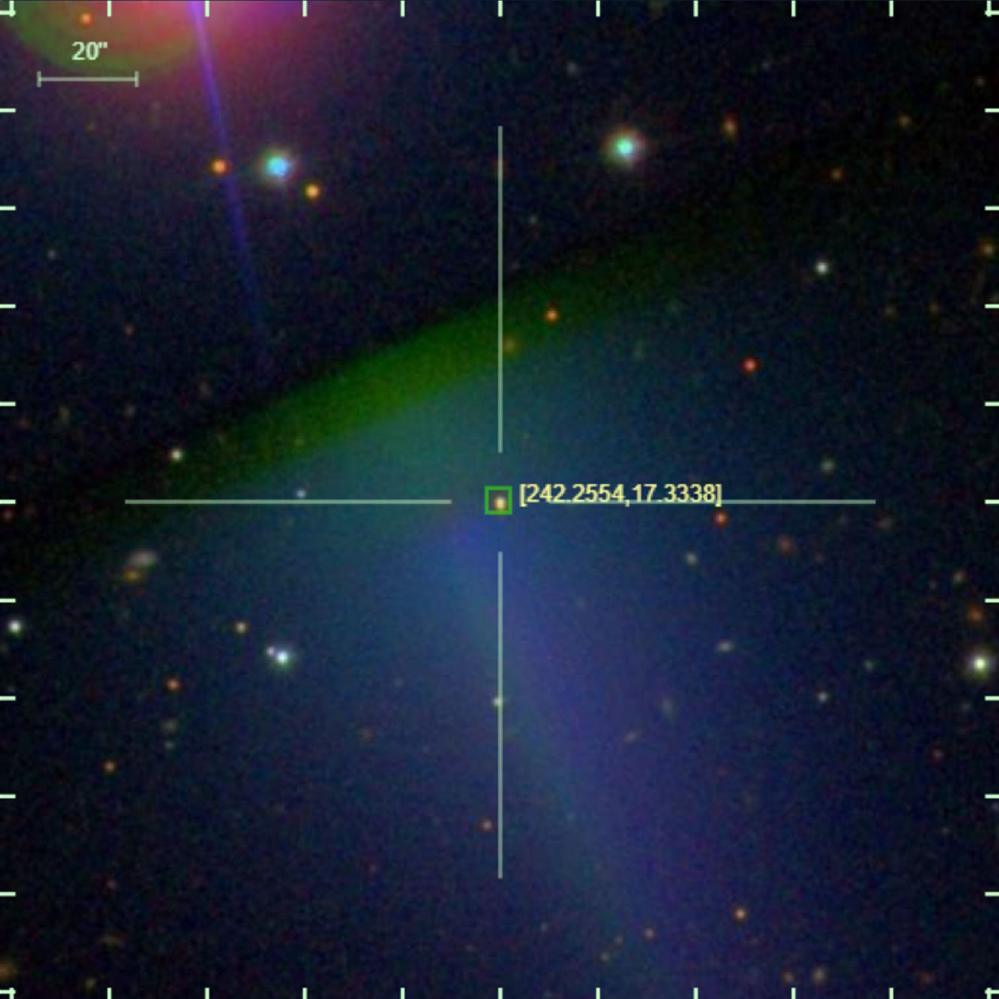}{0.24\textwidth}{(d)}}
\caption{Examples of SDSS imaging of NSA contaminant objects which are removed from our catalog.
(a) Imaging shows a bright, star forming region of a galaxy (NSAID: 177945). (b) A foreground star centered on a distant galaxy which causes incorrect redshift values for the galaxy(NSAID: 401970). (c) An imaging artifact treated as a galaxy by NSA (NSAID: 460798). (d) An artifact from a bright star in the top left surrounds a nearby star in the center, which causes both classification as a galaxy and provides redshift of the star (NSAID: 460798).
\label{fig:badnsa}}
\end{figure*}

HyperLeda gives galaxy diameter measurements as the major axis using a standard $25~$mag~arcsec$^{-2}$ isophote ($d25$), whereas LVG uses a diameter measurement corresponding to the Holmsberg isophote ($\sim26.5~$mag~arcsec$^{-2}$; which we call $d26$). We compared $d25$ HyperLeda diameters to $d26$ LVG diameters for galaxies found in both source catalogs to find the typical ratio between these, and then multiplied the LVG diameters by $\sim0.838$ to approximate $d25$ HyperLeda values for all the nearest neighbor galaxies.

For each NSA object, we then divided the angular separation from its nearest neighbor's angular diameter to create a "match separation ratio", $\delta_{sep}/d25$. Of the galaxies confirmed to include in our sample, we only included those with a ratio $\delta\geq 0.75$ (i.e.~lying beyond 1.5 galaxy radii). Of the 208 objects in category 4, 170 galaxies passed this cut and were added to our final sample. These galaxies have a median distance of 29.9~Mpc and a median mass of $8.08\times10^{7}$~M$_\odot$ ($log(M)\sim$8), and thus add significantly to our sample.

We note that many studies \citep[e.g.][]{geha12,reines13} have used the NSA catalog as a sample of nearby galaxies without removing contaminants.  We find that $\sim$13\% of nearby NSA galaxies (1106/8242) are contaminants. Most contaminants show a low mass estimate from NSA redshift distances and photometry, with a median mass of $2.6\times10^{7}$~M$_\odot$. We therefore include the full list of NSA objects that don't match HyperLeda \& LVG and the criteria we used to cut these down in Table~\ref{tab:nsaflags} for future users.

\subsubsection{Galaxy Coordinates}
\label{subsec:coordinates}
We want the coordinates in our catalog to be centered as accurately as possible on galaxies' nuclear regions for the purpose of AGN selection and observation. Our catalog combines multiple subsamples, each with slightly different astrometric methods and sources. HyperLeda applies a homogenized position computed by averaging published measurements weighted by an assigned quality flag. NSA provides coordinates of the measured object's photometric center in the SDSS. The source of LVG coordinates was not given.  Each time we combined subsamples with different astrometric sources, we compared the coordinates for galaxies in both. We visually inspected 150 galaxies with overlap between HyperLeda and either LVG (739 galaxies) or NSA (7136 galaxies). We randomly selected these galaxies to have a range of offsets between 1\arcsec~ and 10\arcsec. We compared the galaxy positions to imaging with SDSS and 2MASS, both of which have good astrometry \citep{pier03,skrutskie06}. This visual inspection suggests that in general the HyperLeda positions are better aligned with the photocenter than the LVG and NSA samples. We therefore use the following priority order for the {\tt ra} and {\tt dec} in our catalog — HyperLeda, LVG, and NSA.

In addition to the comparison and selection between position sources from multiple catalog sources, we also compared our selected Right Ascension and Declination values to the primary position provided by NED. We retrieved NED positions using a name-based query.  After removing 25 matches with separation $>$5\arcmin~ due to misnamed entries in HyperLeda and NED, we reliably matched positions for 14412 catalog galaxies ($\sim93\%$). We used the visual inspection process described in the previous paragraph to determine whether NED coordinates are noticeably preferable to our catalog position.  We found our positions were generally better than those from NED, with a larger spread between our catalog positions and NED than between HyperLeda, LVG, and NSA.  We find 1680 galaxies have a separation $>3\arcsec$ between our catalog position and the NED position. We investigate the relation between coordinate offsets, mass, and galaxy type in more detail in \S\ref{sec:xray}. Because of the variation between positions from different sources, we choose to include columns for NSA and NED positions in the final catalog, in addition to our chosen compilation of coordinates described above.

\begin{deluxetable*}{ccccr}
\caption{NSA objects that don't match galaxies in the Hyperleda and LVG catalogs.}
\tablehead{\colhead{NSAID} & \colhead{RA} & \colhead{DEC} & \colhead{Quality\_FLAG} & \colhead{MATCH\_RATIO}}
\decimals
\startdata
30129 & 343.8028213 & -0.350540038 & 2 & 191.944 \\
34438 & 13.23646298 & -1.174127269 & 2 & 225.389 \\
36914 & 24.69818421 & 1.071791959 & 2 & 371.004 \\
38419 & 32.6382837 & 0.738166461 & 4 & nan \\
41803 & 8.850357094 & 14.79454207 & 2 & 122.785 \\
42629 & 14.93267592 & 14.56776754 & 2 & 331.128 \\
46179 & 116.1573335 & 40.88379846 & 4 & 151.729 \\
50066 & 129.4311915 & 51.64175054 & 1 & 0.752 \\
51666 & 134.7746823 & 53.76023531 & 1 & 0.707 \\
52302 & 138.5671796 & 57.04441951 & 4 & 318.963 \\
\enddata
  
%Stub of unique NSA objects we individually investigated. NSAID: unique ID within NSA catalog. MATCH\_RATIO: ratio of separation distance and major axis diameter of nearest sample match. QUALITY\_FLAG: integer flag we assigned to differentiate between contaminants and NSA objects to include in the sample.}

\end{deluxetable*}
 \label{tab:nsaflags}

\subsection{Color and Luminosity} \label{subsec:color}

The luminosities and colors of our galaxies are two crucial parts of our catalog.  In addition to being intrinsically useful, the colors are also important for calculating stellar masses (\S\ref{sec:mass}) and for assessing the galaxy type (early vs.~late; \S\ref{sec:type}).  To obtain these estimates, we use four different sources for colors and associated luminosities: (1) HyperLeda, (2) NSA, (3) the Siena Galaxy Atlas (SGA), (4) NASA/IPAC Extragalactic Database (NED). All luminosity calculations utilize the best distances described in \ref{sec:distance}.

\textit{Extinction Correction}: For our foreground extinction corrections, we uniformly apply the \citet{schlafly11} corrections; all our sources have \citet{schlegel98} extinction estimates available, and we multiply these by 0.86 to convert them to \citet{schlafly11} values\footnote{\href{https://irsa.ipac.caltech.edu/applications/DUST/}{\tt https://irsa.ipac.caltech.edu/applications/DUST/}}. For HyperLeda, we used $B$ and $V$ magnitudes, and scaled their $A_B$ as described above; then to determine $A_V$ values we assumed an $A_V/A_B=0.769$ \citep{cardelli89}. For NSA, we used the provided \citet{schlegel98} extinction values for each band and scale these by 0.86.  We retrieved a table of location-based extinction values for our entire catalog from the IPAC Galactic DUST web service and added the table's \citet{schlafly11} $E(B-V)$ values to the catalog as {\tt EBV\_irsa}. For SGA, we then created band-specific extinction arrays by multiplying {\tt EBV\_irsa} by the proper $A/E(B-V)$ values: $g$=3.303, $r$=2.285.  Finally, for NED, we use $B$ and $R$ band magnitudes from the same source for consistency -- these primarily come from the southern ESO Uppsala survey \citep{lauberts89}; we use the HyperLeda B-band extinction, and assume an $A_R/A_B = 0.628$ based on the extinction curves of \citet{cardelli89} and \citet{odonnell94} taken from the Padova CMD website \footnote{\href{http://stev.oapd.inaf.it/cgi-bin/cmd\_3.4}{\tt http://stev.oapd.inaf.it/cgi-bin/cmd\_3.4}}.
 
 We did not apply any internal extinction corrections, thus the luminosities and colors presented here are only foreground extinction corrected.  Internal extinction values are available from both HyperLeda and LVG.  However, we found no correlation between these two.\\

\textit{Luminosities \& Colors from HyperLeda ($(B-V)_0$, $L_V$, $L_B$)}: HyperLeda provides two different $B-V$ colors: (1) the total asymptotic corrected color, {\tt bvtc} which is corrected for both internal and foreground extinction, and (2) color measured within the effective aperture in which half the total B-flux is emitted, {\tt bve}, which is not corrected for extinction. Various HyperLeda galaxies have one or both measurements available. Our first step to obtain consistent values was to undo the HyperLeda extinction corrections on the {\tt bvtc} column, and apply the \citet{schlafly11} foreground extinction correction to both the total and effective colors -- we call these extinction corrected quantities $(B-V)_{t,0}$ and $(B-V)_{e,0}$ respectively.  Because we are using the colors to calculate total masses, we also attempt to correct the small offset between the  $(B-V)_{t,0}$ and $(B-V)_{e,0}$ -- the difference between these colors for galaxies that have both values are shown in Fig.~\ref{fig:bvcolor}. The offset is clearly color dependent, and the line in that figure shows the color dependent correction appslied to the $(B-V)_{e,0}$ to get our final $(B-V)_0$ values for galaxies that only had {\tt bve} data available.  We flagged these data using the {\tt bvcolor\_f} value in our table.  

For calculating luminosities and total magnitudes in $B$ and $V$ bands, we found that the given {\tt bt} and {\tt vt} columns were not consistent with the given colors (after correcting for the different extinction corrections applied) -- this is likely due to the heterogeneity of data sources used. Given the ubiquity of {\tt btc} values, we first calculate the $B$ total magnitude by removing the extinction corrections and applying \citet{schlafly11} foreground extinction corrections to get total $B_{t,0}$ values. We then derived the total V band magnitude by using our $B_{t,0}$ and subtracted the $(B-V)_{t,0}$ value to get a foreground extinction $V$ band magnitude, $V_{t,0}$. We note the use of $B_{t,0}$ values throughout this paper when plotting galaxy luminosities. These are transformed to luminosities using the Vega-based absolute magnitudes of the Sun from {\tt http://mips.as.arizona.edu/\\~cnaw/sun.html}. \\

\textit{Luminosities \& Colors from NSA ($(g-i)_0$, $L_i$, $L_B$)}: NSA provides Petrosian flux and extinction values for each $FNugriz$ band. We calculate the Pogson $g$-, $r$- and $i$-band  magnitudes using the SDSS relation\footnote{\href{https://www.sdss.org/dr12/algorithms/magnitudes/}{\tt https://www.sdss.org/dr12/algorithms/magnitudes/}}: $m=22.5-2.5log_{10}f$. We then extinction correct all individual magnitudes as described above.  For calculating galaxy masses (\S\ref{sec:mass}), we use the $g_0$ and $i_0$ magnitudes to calculate an extinction corrected $(g-i)_0$ color.  We also calculate and $i$-band luminosity for the galaxies ($L_i$) using the solar AB magnitude from \href{http://mips.as.arizona.edu/\~cnaw/sun.html}{\tt http://mips.as.arizona.edu/\~cnaw/sun.html}. To analyze a single luminosity distribution for our entire sample, we also estimate the $B$-band luminosity $L_B$ from the SDSS magnitudes using the stellar transformation $B_0 = g_0 + 0.39*(g-r)_0 + 0.21$ from \citet{jester05}. \\

\textit{Luminosities \& Colors from SGA ($(g-r)_0$, $L_r$, $L_B$)}: 
A complementary source of photometry for primarily southern galaxies is the Siena Galaxy Atlas (SGA\footnote{\href{https://www.legacysurvey.org/sga/sga2020/}{\tt https://www.legacysurvey.org/sga/sga2020/}}).  We matched our catalog to the SGA catalog after removing sources with known redshifts above 3500~km~s$^{-1}$ using a maximum separation distance of 30\arcsec. In cases of multiple galaxies within this threshold, we matched the galaxy with the smallest angular separation. This gave us $(g-r)_0$ data for 7681 galaxies, 1616 of which otherwise lacked color.

The SGA catalog contains a number of different magnitude measurements.  To ensure consistency, we compared overlapping sources with NSA and found {\tt [G,R,Z]\_MAG\_SB26} best matched the NSA magnitudes. We then calculated reddenings and extinction using the IRSA online portal\footnote{\href{https://irsa.ipac.caltech.edu/applications/DUST/ }{\tt https://irsa.ipac.caltech.edu/applications/DUST/ }} to get \citet{schlafly11} extinction values. We corrected for these values to calculate the $(g-r)_0$ color, $L_r$, and estimate the $B$-band luminosity $L_B$ using the same \citet{jester05} equation applied to NSA photometry. \\

\textit{Luminosities \& Colors from NED ($(B-R)_0$, $L_R$, $L_B$)}: To expand our mass estimations for southern galaxies, we used astroquery to request NED photometry for each sample object. We added available $B$ and $R$ data to the sample, specifically objects with an observed passband of {\tt B (Cousins) (B\_25)} and {\tt R (Cousins) (R\_25)}. If multiple magnitude measurements were available for a single object, we included a mean value. Queries returned a uniform $B$ and $R$ magnitude error of 0.09 for all objects. We used these $B$ and $R$ band magnitudes, corrected for foreground extinction, to calculate the $(B-R)_0$ color, $L_R$, and $L_B$ where available.

\textit{Combined Color and Luminosity Estimates}:
In section~\ref{sec:type} we describe color transformations between $(g-i)_0$ and other colors. We use these to derive an estimate of the $(g-i)_0$ color for all galaxies with colors. This is contained in column {\tt gi\_color}. For galaxies with multiple available colors, we give precedence to source colors in order: NSA $(g-i)_0$, and then synthesized $(g-i)_0$ values based on $(g-r)_0$, $(B-V)_0$, and $(B-R)_0$. We compile the $L_B$ values from all sources in column {\tt B\_lum}; most of these are taken directly from HyperLeda, while a handful are synthesized from other sources as described above with the precedence order being HyperLeda, NED, NSA, SGA.  A vast majority of $L_B$ estimates come directly from Hyperleda.

\subsection{Cleaning the Photometric Measurements Used for Mass Estimates} \label{subsec:cleaning}

Due to a number of objects with both NSA and HyperLeda data available, we wanted to test which source provided more reliable photometry when both were available. We compared $g$ vs $B$ magnitudes, due to their similar wavelength, and investigated galaxies with significantly differing magnitudes to available NED values. Generally, by comparing to available photometry compiled in NED, we find that NSA photometry is more reliable in cases where it is discrepant from HyperLeda and therefore typically prioritize NSA data in estimating galaxy masses.  An exception to this is for a selection of galaxies with NSA $g>16$ and $B-g>1$; for these galaxies we use Hyperleda values preferentially, as the HyperLeda values more closely matched other measurements from NED.  Based on this analysis, we created the {\tt mag\_flag} column to signify which source provided the more accurate measurement at a given magnitude -- 0: the difference between $B$ and $g$ are within one magnitude, so NSA values are used for calculating masses. 1: $B$ and $g$ are discrepant; NSA photometry is more reliable. 2: $B$ and $g$ are discrepant; HyperLeda photometry is more reliable, and is used for estimating masses.

There are 446 additional objects where the NSA photometry is not used.  These include objects with flags indicating unreliable photometry, specifically those with the 5th or 6th bitwise flag set in the NSA {\tt DFLAGS} field (Blanton, M., {\em private communication}).  We also don't use NSA photometry for 20 objects with $g$ or $i$ petrosian flux $\leq 0 ~	nanomaggies$. Of these 446 galaxies, 306 had other colors and luminosities available for calculating masses.

\begin{figure}
    \plotone{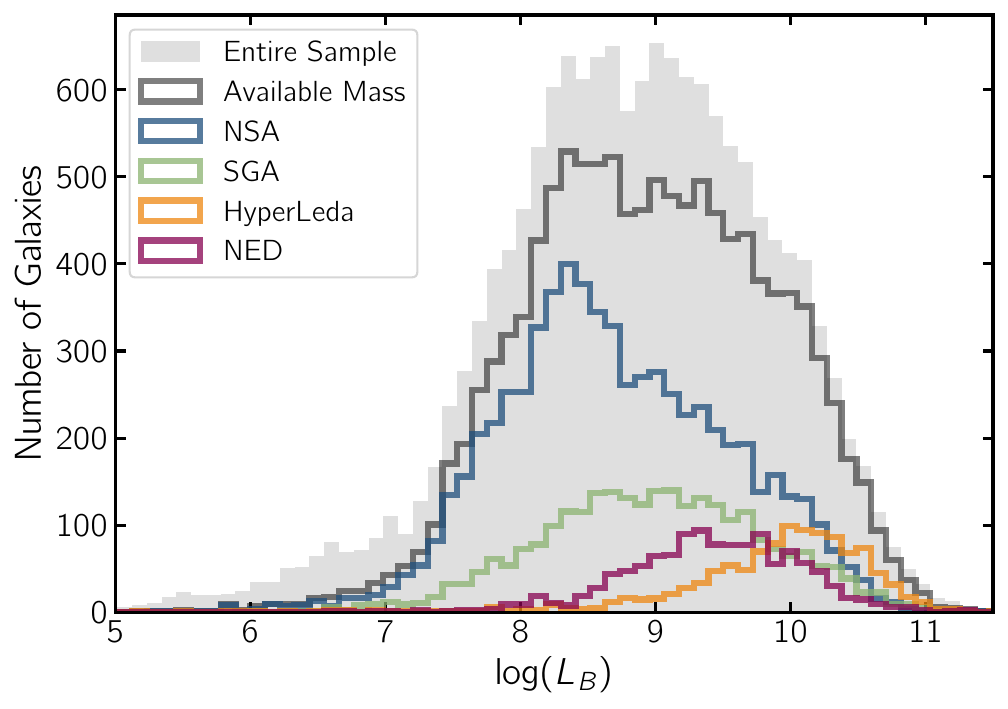}
    \caption{The distribution of the B-band luminosities $log(L_B)$ for the entire catalog (shaded histogram), along with the subset that have available masses (dark gray line).  The colored lines indicate the source of photometry used in estimating the galaxy masses (see \S\ref{sec:mass}.)} \label{fig:loglhist}
\end{figure}

\section{Deriving Best Distances} \label{sec:distance}

In this section we discuss our derivation of the best available distance estimates across our full galaxy sample.  These distance estimates fall into three broad categories: (i) redshift-based distances determined using the Cosmicflows-3 Distance--Velocity Calculator \citep{kourkchi20b}, (ii) redshift-independent distances taken from the LVG and from the NASA/IPAC Extragalactic Database  \citep[NED][]{steer17,steer20}, and (iii) Virgo Cluster distances, for which objects matching Virgo samples and catalogs were assigned either measured distances or an appropriate mean distance.  In general, we prioritize the redshift independent and Virgo distances except in cases where the available redshift based distances are more accurate. $\sim20\%$ of redshift distances are more accurate, once distances reach $>30~$Mpc and error becomes a lower fraction of total distance.

\subsection{Redshift Based Distances}
Due to the peculiar motions of galaxies, in the nearby Universe, redshift based distances are not always accurate.  However, this accuracy can be improved by using models of the local mass density based on the measured peculiar motions of galaxies \citep{courtois12}.  We chose to use the Cosmicflows-3 Distance--Velocity Calculator \citep{kourkchi20b} because it covered the full distance range of our sample using an up-to-date local mass density model \citep{graziani19}.  We entered Heliocentric velocities into the calculator to retrieve distance estimates corrected for the local mass density. Velocity measurements are given precedence in the order NSA, LVG, and HyperLeda where available.

To estimate errors on the redshift based distances, we first estimated a typical peculiar velocity for galaxies in the Local Volume.  We based this on galaxies with redshift independent measurements (see next subsection) with errors $<$10\%.  We then used the CMB-relative velocity measurements from HyperLeda to calculate the peculiar velocity of each galaxy using $ v_{pec} = v_{rad} - H_{0}d$ assuming $H_0 = 70$.  We then measured the width of this distribution using the 16th and 84th percentiles, taking half the width to get a typical peculiar velocity of $\sim$400~km~s$^{-1}$.

Our redshift based distance errors were then calculated by adding and subtracting this typical peculiar velocity from the Heliocentric velocity and recalculating the distance using  the Cosmicflows-3 Distance--Velocity Calculator.  The final redshift-dependent distance error uses a symmetrization of the difference between upper- and lower-limit corrected distances. These distance errors are typically $\sim$65.3\% at 10~Mpc and $\sim$17.4\% at 40~Mpc.

\subsection{Redshift Independent Distances}
We use redshift independent distances from three sources: (1) the updated LVG catalog distances and (2) the compilation of redshift independent distances in the NED distance database \citep[NED-D][]{steer17}.
We chose to give highest priority to the LVG catalog distances, as these distances have typically represent the best available measurements and are kept up to date.  We manually examined distance estimates where the values from LVG conflicted with those NED-D and NED-MED, and found LVG to typically be superior; for instance the distances to the Maffei group galaxies were based on the recently updated TRGB distances presented in \citet{anand19}.

For galaxies in NED-D, we chose our distances using the following list of primary indicators, in order of precedence based on the order given in \citet{steer20}: Cepheids, TRGB, RR Lyrae, Red Clump, SNIa, FGLR, Horizontal Branch, SBF, GCLF, CMD, Type II Cepheids, Miras, PNLF, AGB, Carbon Stars, SNIa SDSS. We use the single most recent preferred distance estimate of the preferred, except if two preferred measurements were from the same year, in which case we apply an average of these preferred distances and errors. If NED-D provided more than two older measurements from the best primary indicator, we ensured that the newest distance was not greater than two standard deviations from the averaged older distances. For galaxies with large spreads in their preferred distances, we use the mean of all preferred distances with the best primary indicator.  We encode this information in the field {\tt 'dist\_nedd\_flag'} - 0: uses single best NED-D measurement, 1: best NED-D distance averaged from multiple measurements, 2: NED-D best indicator mean used, due to discrepancy, as previously described.

\subsection{Galaxies in Virgo}

Due to the nearby Virgo Cluster's large velocity dispersion, redshift based distances to Virgo are much more inaccurate than for galaxies in the rest of our sample.  Because of this, we treated Virgo as a special case for distance purposes.  
We matched and compared our sample to galaxies from \citet{mei07}, the Extended Virgo Cluster Catalog \citep[EVCC;][]{kim14} -- we included "certain members" from their catalog as special cases to have their distances revised. These galaxies have velocities within the range of galaxies gravitationally bound to Virgo at a given projected radius.

For these galaxies, we first combined the separate redshift-based and redshift-independent distance lists, using whichever available option had the lowest fractional error. Galaxies with surface brightness fluctuation distances from from \citet{mei07} are assigned that distance and error. Other Virgo galaxies with only redshift based distances available are assigned the average distance of $16.5 \pm 1.1~$Mpc given by \citet{mei07}. For galaxies with redshift-independent distances, we found more divergent values than expected, including some that didn't agree with Virgo distances. After careful evaluation of individual cases, we assigned the $16.5 \pm 1.1~$Mpc distance for most objects, keeping the redshift independent distance only if  $|D-16.5|>3\sigma$, where $\sigma$ is the error on the distance; this is a total of 862 objects.

\subsection{Final Compilation of Distances}

For galaxies with multiple distance sources, our aim was to identify the more reliable and precise distance source.  Our distance sources included: 
Virgo distances, LVG distances, NED-D distances with lowest error, redshift-based distances from Cosmicflows-3 calculator. We prioritized Virgo distances, and otherwise chose the distances with the lowest error  between NED-D and CF3 z-distances.

Of the 55 objects which had no assigned distance from any of our sources, 29 are assigned the given distances and errors from HyperLeda. The remaining 26 are NSA-only objects for which velocity measurements returned a Cosmicflows-3 corrected distance $D\leq0$; we don't supply a distance in this case.

\begin{figure}
    \plotone{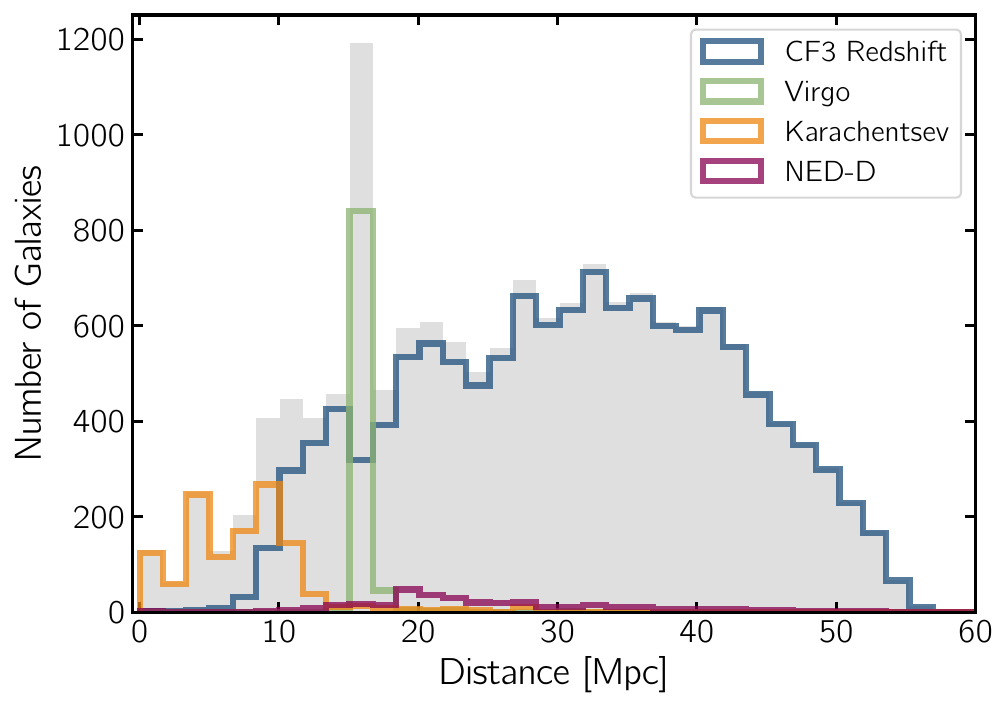}
    \caption{Distribution of our chosen best distances for the entire catalog (shaded histogram). Colored lines indicate the source ({\tt bestdist\_source}) used for the final compilation of distances (see \S\ref{sec:distance}).}
    \label{fig:distances}
\end{figure}

As well as best distances (the {\tt bestdist} column) and errors ({\tt bestdist\_error}), each galaxy in the catalog has a field for specific distance indicator if available. We have also included the field {\tt bestdist\_source}, the values of this are --  0: Virgo, 1: LVG, 2: NED-D, 3: redshift-based, 4: HyperLeda, ‘nan’:No Distance. The final distribution of galaxy distances, separated using these classifications, is shown in \S\ref{fig:distances}. 

\section{Mass Estimation} \label{sec:mass}

Our goal in this section is to estimate consistent stellar masses across the full sample based on galaxy colors and luminosities.  We use the $g-i$ vs. $M/L_i$ color-M/L relation from \citet{taylor11} as the basis for our measurements and derive self-consistent relations in other colors using galaxies with overlapping measurements.

We use the $g-i$ vs.~$M/L_i$ relation as our preferred relation as previous work has found it is accurate to $\sim$0.1 dex due to the alignment of the extinction and stellar population effects \citep{zibetti09, taylor11}.  We note that while NIR photometry is widely used to estimate stellar masses assuming a constant $M/L$ \citep[e.g.][]{meidt14}, this approach is compromised in galaxies with younger stellar populations by its sensitivity to difficult to model TP-AGB stars and hot dust emission \citep[e.g.][]{kriek10,taylor11,mcgaugh14,telford20}.

One challenge in deriving stellar masses in our sample is the different photometry available for different galaxies.  This creates two issues, first,  \citet{mcgaugh14} has shown that individual color-M/L relations are not consistent across multiple bands, i.e. that using different colors in the \citet{bell03} or \citet{zibetti09} relations gives systematically different masses for observed galaxy colors.  Second, many available relations are not available across both the Sloan and Johnson photometry that we are using for our galaxy sample \citep[e.g.][]{roediger15}.  To solve both of these issues, we therefore translate the \citet{taylor11} $g-i$ vs. $M/L_{i}$ into the other colors used ($(g-r)_0$, $(B-V)_0$, and $(B-R)_0$) using overlapping galaxies to ensure self-consistency.

We first used Equation (7) from \citet{taylor11}:
\begin{equation}
  {\rm log}(M_{\star,i}/L_{i})=-0.68+0.70(g-i)
  \label{eq:ML_i}
\end{equation} 
to calculate galaxy stellar mass estimates (log($M_{\star,i}/L_{i}$)) using $(g-i)_0$ and $L_{i}$ derived from NSA data. To translate this relation into other bands, we used overlapping samples. First, for 6677 NSA galaxies with both $(g-i)_0$ and $(g-r)_0$ color available, we compared the stellar masses calculated from $i-$band to the $r-$band luminosities, i.e. log($M_{\star,i} / L_{r}$). We then use this to fit the $M_{\star,r}/L_r$ vs. $(g-r)_0$ relationship using the {\tt RANSAC} fitting from {\tt sklearn.linear\_model} to create a linear fit to this comparison while ignoring outliers as shown in Fig.~\ref{fig:ml_transform}. We followed the same process with 986 overlapping galaxies between NSA and HyperLeda to translate the color--M/L relation to $(B-V)_0$. Our sample lacks significant overlap between galaxies with both $(g-i)_0$ and $(B-R)_0$, due to the latter coming primarily from southern sky observations. We therefore used the same method on 822 galaxies with $(g-r)_0$ and $(B-R)_0$ to bootstrap an appropriate and self-consistent $M/L$ relation. The resulting equations from these transformations are:
\begin{equation}
    {\rm log}(M_{\star,r}/L_{r}) = -0.66 + 1.20(g-r)_0
    \label{eq:ML_r}
\end{equation}
\begin{equation}
    {\rm log}(M_{\star,V}/L_{V}) = -0.72 + 1.07(B-V)_0
    \label{eq:ML_V}
\end{equation}
\begin{equation}
    {\rm log}(M_{\star,R}/L_{R}) = -0.51 + 0.51(B-R)_0
    \label{eq:ML_R}
\end{equation}

We multiplied the color-based M/L ratios by their corresponding luminosity to calculate a stellar mass estimate. We created separate columns for mass estimates from each color: $g-i$, $g-r$, $B-V$, and $B-R$. To compile our final column of "best" mass estimates for 11740 galaxies, we gave precedence when available to (1) 6626 log($M_{\star,i})$ values calculated using NSA data using Equation~\ref{eq:ML_i}, (2) 2705 log($M_{\star,r})$ values from SGA data using Eq.~\ref{eq:ML_r}, (3) 1182 log($M_{\star,V})$ values from HyperLeda using Eq.~\ref{eq:ML_V}, (4) 1227 log($M_{\star,R})$ values from retrieved NED photometry using Eq.~\ref{eq:ML_R}.

To obtain errors on the mass estimates, we use the scatter from multiple estimates for the same objects.  To do this, we take the sample of 986 galaxies with both $(B-V)_0$ and $(g-i)_0$ colors, and compare the masses derived with these two separate colors.  This comparison includes scatter in the mass estimates due to different bandpass information and independent photometric measurements. The 68 percentile of the mass differences is 0.136. We use this value as an assumed floor on the log($M_{\star}/M_{\odot}$) error. We also propagated distance errors by calculating high and low masses adding and subtracting {\tt bestdist\_error}. For our final {\tt logmass\_error}, we use the larger of these two error estimates.

\subsection{Mass Comparisons}
\label{subsec:masscomparison}

Our mass estimates are based on those of \citet{taylor11} and thus should be consistent with the widely used GAMA mass functions \citep{baldry12,kelvin14,wright17,driver22} that provide the deepest mass function estimates for galaxies in the local Universe.  The \citet{taylor11} mass estimates are based on the \citet{bruzual03} stellar population models and assume a \citet{chabrier03} mass function and the dust extinction law from \citet{calzetti00}.  There is an extensive comparison of how this color-M/L relation compares to earlier color-M/L relations in \citet{taylor11}. In particular, their Fig.~13 shows a their relation is $\sim$0.2 dex below the \citet{bell03} relation (due to IMF differences) and a similar normalization but significantly different slopes from the \citet{zibetti09} relationship due to differences in the weighting of stellar population models.  Here we compare our derived masses against dynamical mass estimates from ATLAS\textsuperscript{3D} \citet{cappellari11} and IR based mass estimates from the S4G survey \citep{sheth10}.  
%ACS pick up here.  

\subsubsection{ATLAS\textsuperscript{3D} Dynamical Mass Comparison}

ATLAS\textsuperscript{3D} is a volume-limited sample of 260 nearby early-type galaxies \citet{cappellari11} within 42 Mpc.  Dynamical masses were derived using Jeans anisotropic modeling of the integral field kinematics in \citet{cappellari13a} and \citet{cappellari13b}.  We compare our masses to those derived from the stellar-only dynamical (M/L)$_{\rm stars}$ estimates from \citet{cappellari13b},  shown in Fig.~\ref{fig:mass_comp}.  The ATLAS\textsuperscript{3D} masses are mostly larger than our masses, with a median offset of \~0.34 dex.  The offset is near zero in the lowest mass galaxies (log($M_\star/M_\odot$)$<$10), and is highest (nearly 0.5 dex) at the highest masses.  These offsets are consistent with the findings of \citet{cappellari12}, who found that a vast majority of ATLAS\textsuperscript{3D} galaxies had dynamical mass estimates that were heavier than stellar population model estimates based on a \citet{chabrier03} IMF, with a typical offset roughly a factor of 2 (0.3 dex) higher.  They found that this offset strongly depends on the observed velocity dispersion, with the highest dispersion and mass galaxies having the largest offsets, just as observed here.  So overall, this comparison suggests that our mass estimates are accurate for their assumption of a Chabrier IMF, but that assumption may be a poor one especially for the most massive early-type galaxies, and our stellar mass estimates may significantly underestimate the masses of these galaxies.

\subsubsection{S4G IR-based Mass Comparison}

\begin{figure}
    \plotone{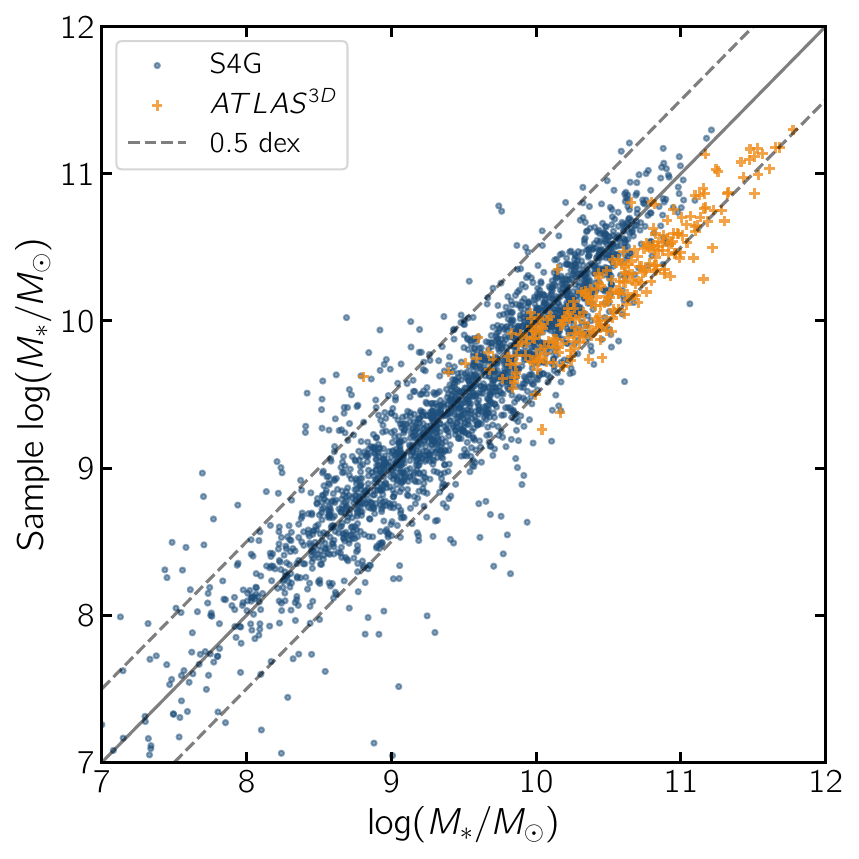}
    \caption{Comparison of our mass estimates ({\tt logmass\_best}) against those from S4G \citep[blue;][]{sheth10, eskew12} and ATLAS\textsuperscript{3D} dynamical stellar mass estimates \citep[red;][]{cappellari11}.  The S4G masses are corrected from a Salpeter to Chabrier IMF by subtracting 0.24 dex for consistent comparison to our catalog.  The solid line shows the one-to-one relation, while dashed lines indicate an 0.5 dex offset.  
    }
    \label{fig:mass_comp}
\end{figure}

The Spitzer Survey of Stellar Structure in Galaxies (S4G) is a volume-, magnitude-, and size-limited survey of over 2300 nearby galaxies at 3.6 and 4.5 \textmu m \citep{sheth10}. The available mass estimates are based on IR luminosities using the relation of \citet{eskew12}.  This relation uses the resolved star stellar mass estimates assuming a Salpeter IMF of the LMC from \citet{harris09} and combines this with IR fluxes 3.6 and 4.5 $\mu$m to get a luminosity to stellar-mass conversion.  This relation was applied to all the S4G galaxies to derive stellar masses available in the v.~2 S4G catalog\footnote{Available at \href{https://vizier.u-strasbg.fr/viz-bin/VizieR?-source=J/PASP/122/1397/s4g}{\tt https://vizier.u-strasbg.fr/viz-bin/\\VizieR?-source=J/PASP/122/1397/s4g}}.  Using these values, we match 
2346/2352 galaxies (99.7\%) and plot our derived masses vs. the S4G \citet{eskew12} masses \citep[after subtracting 0.24 dex to account for the difference expected in mass between a Chabrier and Salpeter IMF][]{cappellari12} in Fig.\ref{fig:mass_comp}.  After the correction for the Salpeter IMF, we see very close agreement between our masses with just a bias of just 0.03 dex (with S4G masses being very slightly more massive). The one-to-one relation extends over a wide range of galaxy masses, and the 1$\sigma$ scatter around this relation is just 0.24 dex.  Overall, we find good agreement between the S4G masses and the masses we derive after accounting for different IMF assumptions.

\subsection{Mass Completeness}
\label{subsec:completeness}

Many science applications benefit from complete, volume-limited samples of galaxies.  To gauge the completeness of our sample, we compare our galaxy catalog to mock galaxy samples created from galaxy stellar mass functions.  Above, we have derived masses for our galaxies based on the color--stellar mass relations used in the GAMA survey \citep{baldry10,taylor11}; we therefore use the GAMA galaxy stellar mass functions to evaluate the completeness of our catalog.

\begin{figure}[!ht]
    \plotone{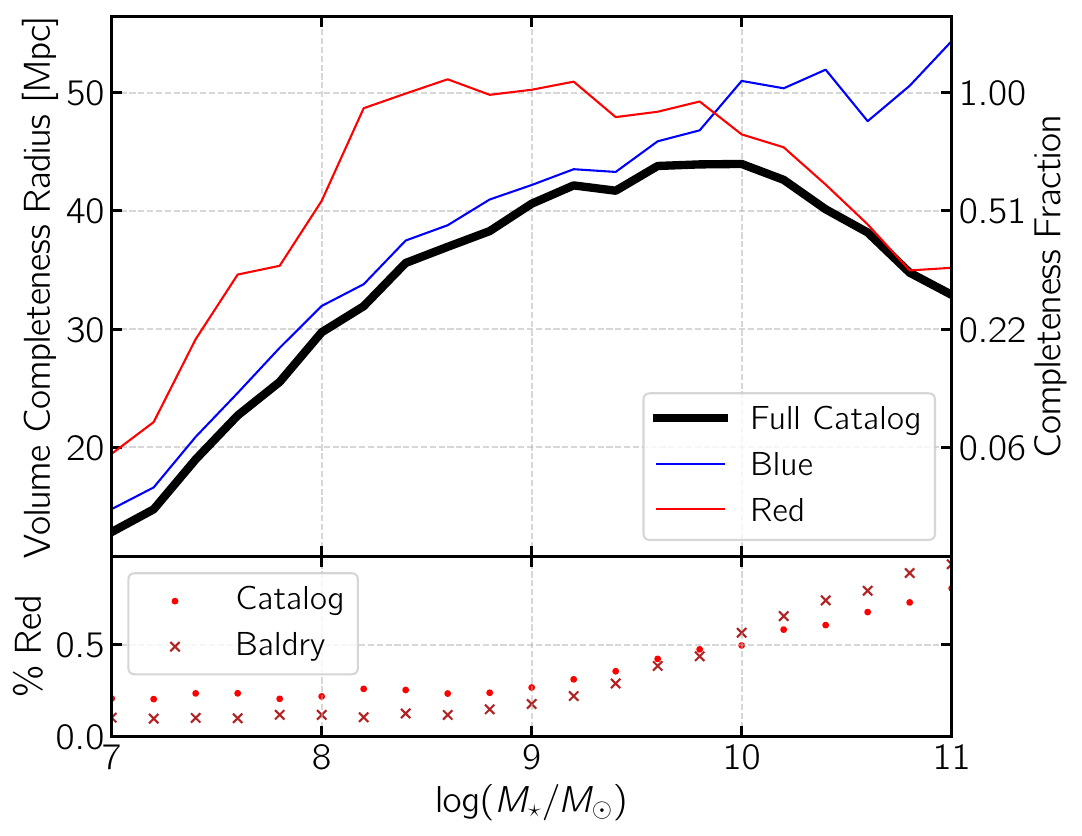}
    \caption{Completeness of our catalog as a function of stellar mass. {\em Top--} Mass completeness of our catalog, using 0.2 dex bins. Y-axes show the radius of effective volume {\em left} and completeness fraction {\em right}. The black line compares all catalog galaxies against a $<50$~Mpc mock sample created using the \citet{driver22} galaxy stellar mass function. Colored lines show comparisons between {\tt color\_type} separated catalog galaxies against color-separated mock sample created using the \citet{baldry12} galaxy stellar mass functions. {\em Bottom--} Percent of galaxies classified as "Red" by our catalog and the \citet{baldry12} sample in each bin.
    }
    \label{fig:completeness}
\end{figure}

\begin{figure}
    \plotone{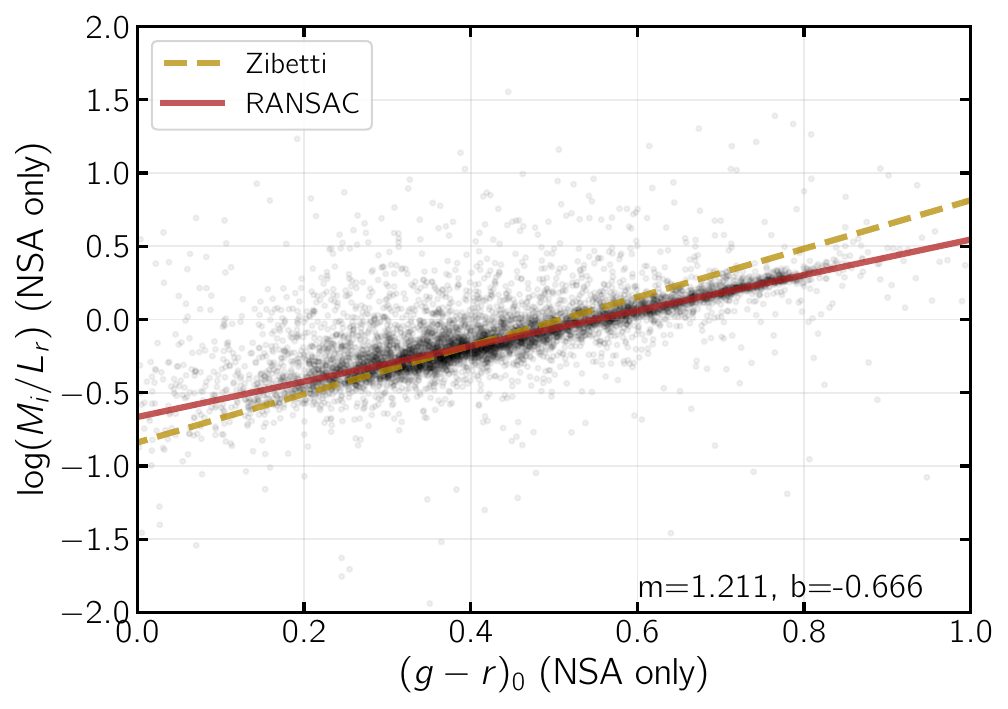}
    \caption{An example of how we transform the \citet{taylor11} $(g-i)_0$--$M/L_i$ relation to $(g-r)_0$--$M/L_r$.  The y-axis is determined by taking masses calculated from the NSA data using the $(g-i)_0$--$M/L_i$ relation, and then dividing by the NSA $L_r$ values.  This is then plotted against NSA $(g-r)_0$ values and fit using a robust outlier-rejection algorithm, {\tt RANSAC}.  Also shown is the somewhat steeper relation from \citet{zibetti09}.  Our translation of the \citet{taylor11} relation to other colors to ensures consistency between masses calculated in different bands.  We note that although this $(g-r)_0$--$M/L_r$ relation is used for galaxies with SGA data, we fit this relation with just NSA photometry to ensure a large sample of galaxies and minimize photometric outliers. A similar process was used to also create the other color-$M/L$ relations given in \S\ref{sec:mass}.}
    
    \label{fig:ml_transform}
\end{figure}

For the full sample of galaxies, we use the most recent stellar mass function published by GAMA from \citet{driver22}, which extends down to $M_{\star}\sim10^{7}M_{\odot}$.  We simulate a galaxy population with a volume of 4/3$\pi$(50~Mpc)$^3$ based on the \citet{driver22} mass function. 
We then compare the number of galaxies with available masses in our catalog to this mock galaxy sample in 0.2 dex mass bins to calculate the fraction of the expected number of galaxies that we detect.  We then translate this fraction to estimate the "Volume Completeness Radius", which we define as the radius out to which we'd have a complete sample based on the total number of galaxies in our sample.  For instance, if we detect ~50\% of the number of expected galaxies, this translates to a Volume Completeness Radius of $\sim$40~Mpc. Fig.~\ref{fig:completeness} shows that the catalog is $>$50\% complete at $M_{\star}\gtrsim10^{8.5}M_{\odot}$, which corresponds to an effective volume with radius $R\sim40$~Mpc.  At lower masses, the completeness drops significantly.  At the highest masses, the sample also appears to become less complete, a point which we come back to below.

The \citet{driver22} paper doesn't have comparable morphology or color separation to what we have done here, so we use an earlier GAMA paper by \citet{baldry12}, which is more directly comparable to our color-based morphology separation (i.e. the {\tt color\_type} in our sample).  We note that the color-separation used here and in the \citet{baldry12} are similar but not identical. Fig.~\ref{fig:completeness} shows that for the late-type galaxies, the Volume Completeness Radius rises to the expected 50~Mpc at stellar masses of $\sim$10$^{10}$~M$_\odot$.  
 The early-types achieve a similar completeness at much lower masses, but then the completeness declines at the highest masses. Values greater than one in this figure could be due to: (1) a real difference in the number density of galaxies in the local Universe than in the regions used for the \citet{baldry12} mass function, or (2) differences in the color separation between our study and theirs.  Note that a direct comparison using identical color separations isn't possible due to their use of $u-r$ colors in separating their populations, and the lack of $u-r$ colors for many galaxies in our sample.
The bottom panel of Fig.~\ref{fig:completeness} shows the fraction of early-type galaxies as a function of stellar mass in our sample compared to the \citet{baldry12} mass function. This shows that the fraction of red galaxies is higher than expected in the local Universe at low masses, but lower at higher masses.

The lack of high mass galaxies overall, and specifically high mass early-type galaxies is a surprising result, because we expect to be complete at the highest masses.  This therefore implies that these galaxies are underdense in the local Universe relative to the volumes probed in the GAMA survey. We re-emphasize that our stellar mass estimates should be consistent with GAMA, thus making the difference more likely to represent a real deficit of massive early-type galaxies locally relative to the GAMA volume.

\begin{figure}
    \plotone{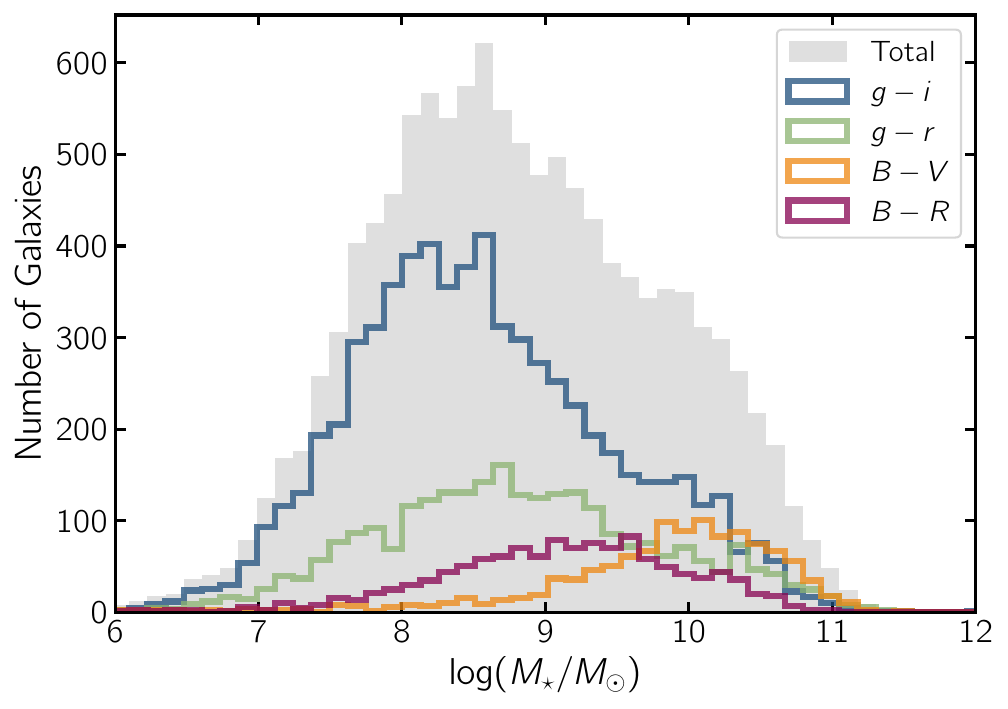}
    \caption{Distribution of log($M_{\star}/M_{\odot}$) estimates for the entire catalog (shaded histogram). Colored lines indicate which colors are used to estimate stellar mass, using color -- mass-to-light ratio relations and precedence as described in \S\ref{sec:mass}.}
    \label{fig:massdist}
\end{figure}

\section{Galaxy Morphologies \& Group Membership} \label{sec:type}

\subsection{Galaxy Morphologies}
\label{subsec:morph}
\begin{figure*}
\plottwo{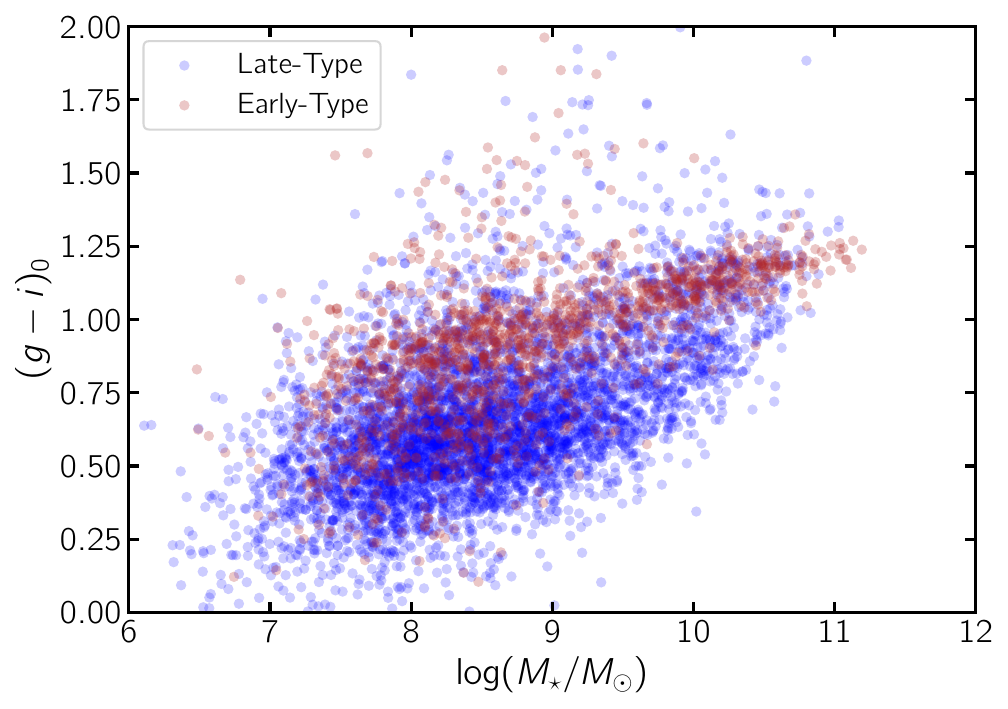}{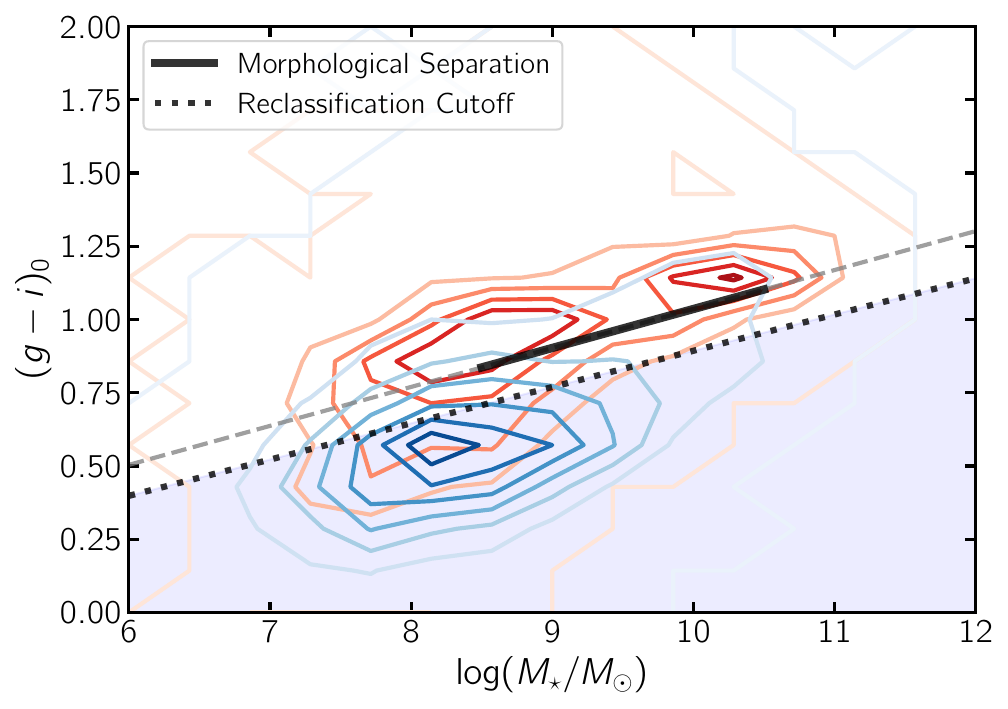}
\caption{Morphological classifications of our galaxies in the color--mass plane.  {\em Left --}  a color--mass diagram of 6677 galaxies with $g-i$ data from NSA and morphologies from HyperLeda and LVG. {\em Right --}  the same color--mass distribution (early-types in red contours, late-types in blue), with the black line showing our linear fit to  most cleanly separate early- and late-type galaxies; this fit is extrapolated to larger and smaller masses (gray line). The dotted line and blue shaded regions shows where we reclassify blue early-type galaxies as late-type based on visual inspection of these galaxies (see  \S\ref{sec:type}). }
\label{fig:colormass}
\end{figure*}

Many galaxies in the nearby universe have been morphologically classified using the Hubble and de Vaucouleurs classification systems.  Galaxies can be separated into two broad categories: early-types are bulge-dominated, elliptical or lenticular in shape, and host little-to-no current star formation. Late-types have a spiral disk shape with smaller central bulges and undergo current star formation. Numerical Hubble T classification ranges from -6 to 10, with negative values corresponding to early-type galaxies and positive values to late-types. In addition, early- and late-type galaxies are often separated using galaxy color-magnitude or color-mass diagrams.  At a given stellar mass, early-type galaxies are typically associated with redder colors, while late-types are associated with bluer colors. In this section, we discuss the source of our morphological type information. We then also describe how we translate this into a color-based separation of early and late-type galaxies. We then evaluate the X-ray active fraction as a function stellar mass and morphology/color in \S\ref{sec:xray}. 

We use the numerical $T$-type translation of the morphological classifications in this work.  Both HyperLeda and LVG provide numerical $T$-types for 13744 of the 15611 (89\%) galaxies in our catalog. Of the remaining galaxies, 603 sample galaxies have available color information. Furthermore, as can be in the left panel of Fig.~\ref{fig:colormass}, the early- and late-type galaxies (a.k.a.~the red sequence and blue cloud) show considerable overlap, especially for low-mass ($<$10$^9$~M$_\odot$) galaxies.  In order to give all sample objects an appropriate morphology, we worked to minimize the potential for morphological misclassification and calculated a color-based type for all galaxies with available color information.

To separate galaxies into early- and late-types using the color-mass diagram, we first calculated a morphological separation line by minimizing the misclassification of galaxies with both morphological $T$-types and NSA colors. We separated galaxies as early- and late-type based on their $T$-type (with $T > 0$ being late-type), fit a line to the color-mass diagram that minimizes the number of misclassifications, i.e. late-types redder than the line, and early-types bluer than the line.  To fit this line, we first divided the data into 0.25 dex mass bins and calculated the fraction of misclassifications as a function of $(g-i)_0$ color. We then plotted the $(g-i)_0$ value with minimum misclassification in each bin and found the best-fit line to to establish our color-based morphological separation line.  We note that we only fit the line between log($M_{\star}$/$M_{\odot}$) of 8.5 to 10.5 where galaxy mass estimates are more robust.
This line's equation is:
\begin{equation}
    (g-i)_0 = 0.133 * {\rm log}(M_{\star}/M_{\odot}) - 0.294
\end{equation}
This equation appears as the upper dashed line in the right panel of Fig.~\ref{fig:colormass}.  

Visual inspection of SDSS imaging of the misclassified galaxies using our morphological separation line suggested that while most red late-types were showed distinct signatures of being dusty late-type galaxies (i.e. spiral arms or clumpy regions of active star formation), many blue early-types were in fact not properly classified and also showed these signatures of ongoing star formation.   This was especially true for early-type galaxies that had colors much bluer than our morphological separation line.  To better separate early- and late-type galaxies we decided to reclassify very blue early-type galaxies.  We created a second reclassification cutoff line below which we reclassified these galaxies in our {\tt best\_type} column (as described below) .  To create this line, we sorted the early-type contaminants
by color in 0.25 dex mass bins. For each mass bin, we visually classified these contaminants as early- or late-type and then calculated threshold color such that legitimate misclassification was minimized within the cutoff.  Then, as above, we fit reclassification cutoff line to the colors determined in each bin to get:  
\begin{equation}
    (g-i)_0 = 0.124 * {\rm log}(M_{\star}/M_{\odot}) - 0.344
\end{equation}
This equation appears as the lower dotted line in the right panel of Fig.~\ref{fig:colormass}.

To apply our color-based reclassification across our full sample, we needed to transform our morphological separation and reclassification cutoff lines to $(g-r)_0$, $(B-V)_0$, and $(B-R)_0$. Note that for deriving the $(B-R)_0$ we added $(B-R)_0$ data from \citet{cook14} for galaxies in our sample.  Using overlapping galaxies and using {\tt RANSAC} as described above, we used the following color transformations.
\begin{equation}
    (g-r)_0 = 0.562 * (g-i)_0 + 0.109 
\end{equation}
\begin{equation}
    (B-V)_0 = 0.625 * (g-i)_0 + 0.131
\end{equation}
\begin{equation}
    (B-R)_0 = 0.898 * (g-i)_0 + 0.133
\end{equation}

Applying these color-transformations we get the following morphological separation lines:
\begin{equation}
    (g-r)_0 = 0.075 * {\rm log}(M_{\star}/M_{\odot}) - 0.056
\end{equation}
\begin{equation}
    (B-V)_0 = 0.083 * {\rm log}(M_{\star}/M_{\odot}) - 0.052
\end{equation}
\begin{equation}
    (B-R)_0 = 0.119 * {\rm log}(M_{\star}/M_{\odot}) - 0.131
\end{equation}

and the reclassification lines:
%three equations
\begin{equation}
    (g-r)_0 = 0.070 * {\rm log}(M_{\star}/M_{\odot}) - 0.126
\end{equation}
\begin{equation}
    (B-V)_0 = 0.062 * {\rm log}(M_{\star}/M_{\odot}) - 0.098
\end{equation}
\begin{equation}
    (B-R)_0 = 0.111 * {\rm log}(M_{\star}/M_{\odot}) - 0.242
\end{equation}

For our final catalog, we provide several morphological types.  In addition to our catalog columns for numerical $T$-types and their source, we created two type columns which broadly categorize galaxies into 'early' or 'late'. Our {\tt color\_type} classifies a galaxy as early if its color lies above the morphological separation line, and late if it lies below the line. Our recommended {\tt best\_type} column follows this process:\\
(1) most galaxies are classified based on their $T$-type from HyperLeda or \citet{karachentsev13} with $t\leq0$ = 'early' and $t>0$ = 'late'.\\
(2) The 387 early-type galaxies that have colors bluer than the reclassification cutoff are reclassified as late-type.\\
(3) 603 galaxies without a morphology are assigned their color-based type. We reclassify each galaxy based on the color used to estimate mass in the final catalog. 

\begin{figure*}
\plottwo{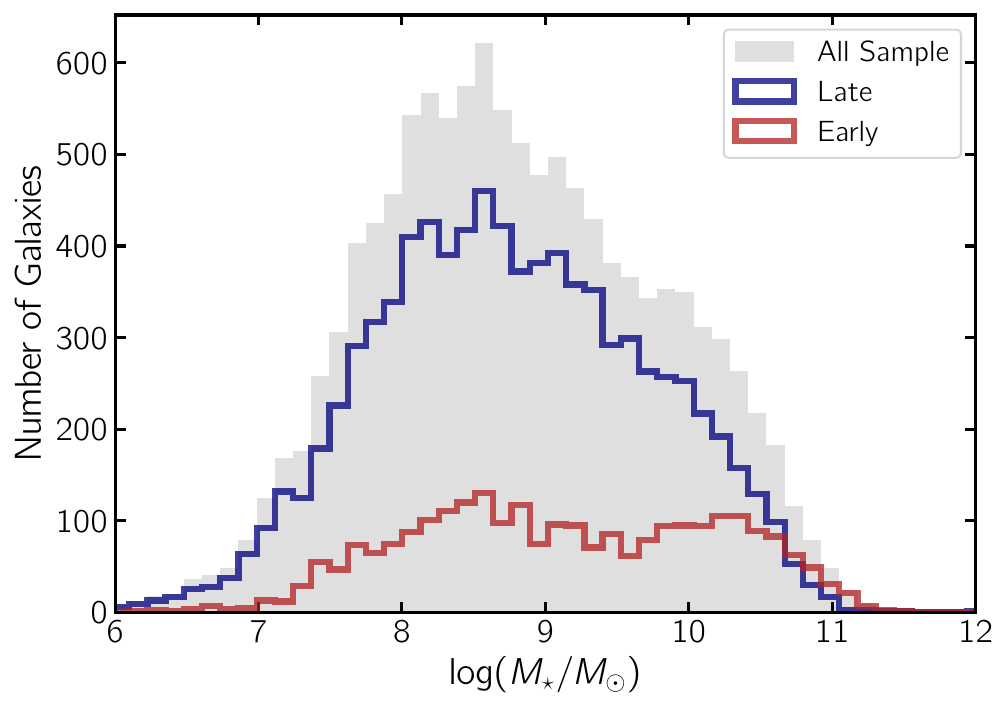}{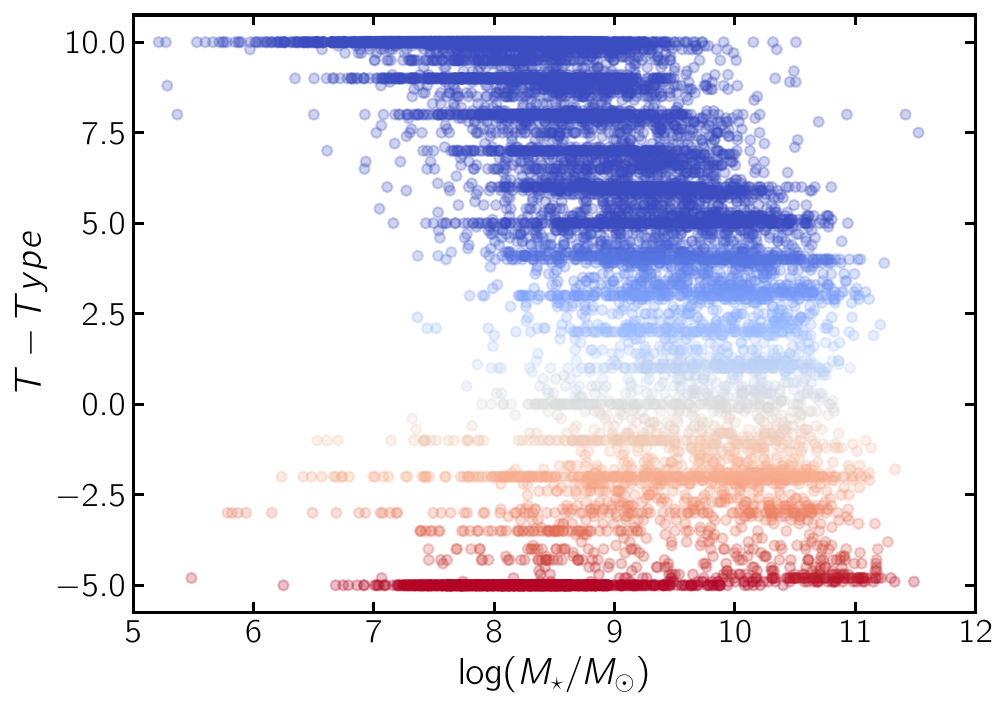}
\caption{Mass distribution of galaxies as a function of their morphology.  {\em Left --} Distribution of log($M_{\star}/M_{\odot}$) for the full catalog (shaded histogram), using {\tt best\_type} as described in \S\ref{subsec:morph} to separate into early- (red) and late-types (blue). {\em Right --} Comparison of numerical T-Type vs log($M_{\star}/M_{\odot}$) for all catalog galaxies with an available {\tt t\_type}. Color gradient is included for visual representation of galaxy type and coincides with T-Type value.}
\label{fig:mass_type}
\end{figure*}

In total, we have {\tt best\_type} values for 14347 out of 15611 galaxies, and {\tt color\_type} values for 11740 galaxies.  We show the masses of galaxies separated into early- and late-type in the left panel of Fig.~\ref{fig:mass_type}, and also include the masses of all galaxies with $T$-types in the right panel. 

\subsection{Galaxy Environments: Group Catalog Assignment}
\label{subsec:groups}

The galaxies in the local Universe span a wide range of environments, from dense clusters like Virgo to galaxies in voids.  One common way of identifying galaxies in dense environments is through group catalogs. To enable easy identification of galaxies in groups in our survey, we use the \citet{lambert20} group catalog based on the 2MRS survey.  This group catalog identifies 3022 groups from the 2MRS catalog (which extends well beyond our 50~Mpc limit) using a friends-of-friends algorithm.  

To match our galaxy catalog, we first used Table 3 from \citet{lambert20}, which gives individual galaxies that are part of a group or subgroup using the paper’s friends-of-friends algorithm. We crossmatched our sample to galaxies with $v_{cmb}<3867$~km~s$^{-1}$ (based on our highest redshift galaxy), resulting in 2463/2864 unique matches within 30\arcsec. We wanted to ensure we also matched low-mass sample galaxies which might not be included in the \citet{lambert20} catalog due to the magnitude limit of 2MRS. To do this, we used Table 1 from \citet{lambert20}, which includes group positions, comoving distances, velocity dispersion, and radius/$R_{200}$ estimates.
We crossmatched our entire sample to find the nearest group for each galaxy, considering sample galaxies to be part of a group if they fulfilled two requirements: (i) the velocity for our sample galaxy must be within 2$\times$ the velocity dispersion of the group, and (ii) the angular separation between between the galaxy and its nearest group matches was smaller than the angular radius of that group, ($d \leq R_{200}$). Using this three-dimensional approximation, we flagged an additional 790 sample galaxies galaxies as likely group members. As expected, the median log($M_{\star}/M_{\odot}$) for the distinct group method matches is $\sim$8.4, compared to $\sim$10 for the direct galaxy matches.

We assigned \citet{lambert20} Group IDs to 3253 sample galaxies, giving precedence to those with direct galaxy matches. This simultaneously flags galaxies as being part of a group, and easily allows users to seek further group information (including group radius and velocity dispersion) by cross-correlating our catalog with the catalogs available from \citet{lambert20} on the publisher's website.  

The Virgo cluster is indicated in the \citet{lambert20} catalog as cluster 2987 -- of the 347 galaxies with that group identification, we find 276 of them (79\%) are included in Virgo based on distance sources in \S\ref{sec:distance}. This is a small fraction of the 907 total galaxies that we identify as being part of Virgo in our catalog because the \citet{lambert20} catalog is based on 2MASS and thus misses many fainter Virgo galaxies.

\section{X-ray Active Fraction Measurement}\label{sec:xray}

\subsection{Background} \label{subsec:xrayintro}
The initial motivation for creating this sample was to quantify the demographics of central black holes in nearby lower mass galaxies.  These demographics can provide constraints on the formation mechanisms of central massive black holes \citep[see recent review by][]{greene20}.  X-ray measurements of AGN in nearby galaxies $\lesssim 50$~Mpc can provide deep and clean enough observations to constrain the occupation fraction of black holes in galaxies below 10$^{10}$~M$_\odot$. Results on large samples of early-type galaxies \citep{miller15,gallo19} suggest a high occupation fraction for 10$^9$ to 10$^{10}$~M$_\odot$ galaxies, in agreement with dynamical measurements in a handful of nearby early-type galaxies \citep{nguyen18,nguyen19}.  However, a majority of galaxies below 10$^{10}$~M$_\odot$ are late-type galaxies \citep{baldry12}, and thus far no suitable galaxy sample exists to compare the X-ray active fraction (and thus occupation fraction) of late-type galaxies to the studies of early-type galaxies at masses below $M_\star \sim $10$^{10}$~M$_\odot$.  Recent compilations of archival X-ray data in nearby galaxy samples cut out most low mass nearby galaxies by requiring redshift independent distance measurements \citep{she17a,she17b}, or by including only brighter samples dominated by massive galaxies \citep{bi20}. The X-ray active fractions of more distant dwarfs (between $z$ of 0.08 and 2.4) have also been studied by \citet{mezcua18} 
 in the COSMOS field. Our new 50 Mpc Galaxy Catalog provides the best starting point for constraining the local X-ray active fraction and the occupation fraction of central black holes. We focus specifically on trying to get as large and complete a sample as possible down to lower masses (i.e. $10^{8} $~$M_{\odot}$), with stellar mass estimates and Hubble type measurements based on color index.
Below we quantify the X-ray active fraction using our new catalog.

Detection of a nuclear X-ray source above $10^{42}~$erg~s$^{-1}$ (the "classical" AGN threshold) is invariably associated with accretion-powered emission from a massive (i.e., non-stellar) black hole. Below this threshold, and even more so below $10^{40}~$erg~s$^{-1}$, emission from bright X-ray binaries could contribute to the detected signal. The expected luminosity from low-mass X-ray binaries in particular scales with stellar mass \citep{lehmer19}, which makes spatial resolution key for discriminating between low-Eddington ratio massive BHs vs. contaminants. Owing to the factor $\sim$4 higher resolution, Chandra is thus preferable to XMM-Newton for searching and identifying massive black holes with luminosities lower than traditional AGN.

We present the Chandra-based X-ray active measurements of our galaxy catalog here, without an attempt to model the distribution or account for contamination from X-ray binaries.  However, in a subsequent paper, we will use this data to constrain the $L_X - M_\star$ relationship and the occupation fraction as a function of galaxy stellar mass including estimates of contamination from X-ray binaries (Gallo et al. {\em in prep}).

\subsection{50 Mpc Active Fraction Measurement}
\label{subsec:activefraction}

We determined which galaxies in our catalog have available Chandra X-ray data using the Chandra Source Catalog 2.0 \citep[][and Evans {\em in prep}]{evans10}, which includes Chandra observations made public before the end of 2014. We search for X-ray sources coincident with our galaxies' nuclei using their catalog  {\tt ra} and {\tt dec}.  Note that our choice of coordinates have been discussed in \S\ref{subsec:coordinates}, and the effect of uncertainties on these central coordinates is considered in more detail in \S\ref{subsec:activesystematics}. Using Chandra's CSCView\footnote{\href{http://cda.cfa.harvard.edu/cscview/}{\tt http://cda.cfa.harvard.edu/cscview/}}, we retrieved two tables from a crossmatch with our catalog: (1) a table of limiting sensitivities for all 1506 objects with Chandra data, and (2) a table of detections within a  given angular separation (1--3\arcsec) of the galaxies' {\tt ra} and {\tt dec}.   The second table of detections contains 291 unique galaxies with matching detections when using a 1\arcsec\ search radius (note that 20 objects returned multiple sources).   We use our distance estimates ({\tt best\_dist}) to calculate X-ray luminosities from the 0.5-7~keV fluxes. These have a median log(L$_{X}$/erg~s$^{-1}$) of 39.21 (hereafter we just refer to this as log(L$_{X}$)), with a full range of log(L$_{X}$) from 33.19--42.26.  While some of these sources are likely contaminants from X-ray binaries and not accreting massive black holes, we attempt to minimize this contamination in two ways: (1) we restrict our X-ray activity analysis below to the 249 (85\%) of sources that have available log($M_{\star}/M_{\odot}$) and log(L$_{X}$)$>$38.3 \citep[as in][]{miller15,gallo19}, and (2) we use a 1\arcsec\ match as our default matching radius.  We note that previous work on X-ray binary contamination in similar Chandra observations has shown that only a small fraction ($\sim$10\%) of nuclear X-ray detections with log(L$_{X}$)$>$38.3 are likely to be contaminants \citep[e.g.][]{gallo10,miller12,foord17}.  We include the columns  {\tt chandra\_observation} and {\tt chandra\_detection} to flag matched catalog galaxies using our default 1\arcsec\ matching, and also include a {\tt chandra\_detection\_3arcsec} for galaxies with sources within 3\arcsec\ (see \S\ref{subsec:activesystematics}); the X-ray luminosity is given in the {\tt log\_lx} column.

\begin{figure*}[!ht]
    \plotone{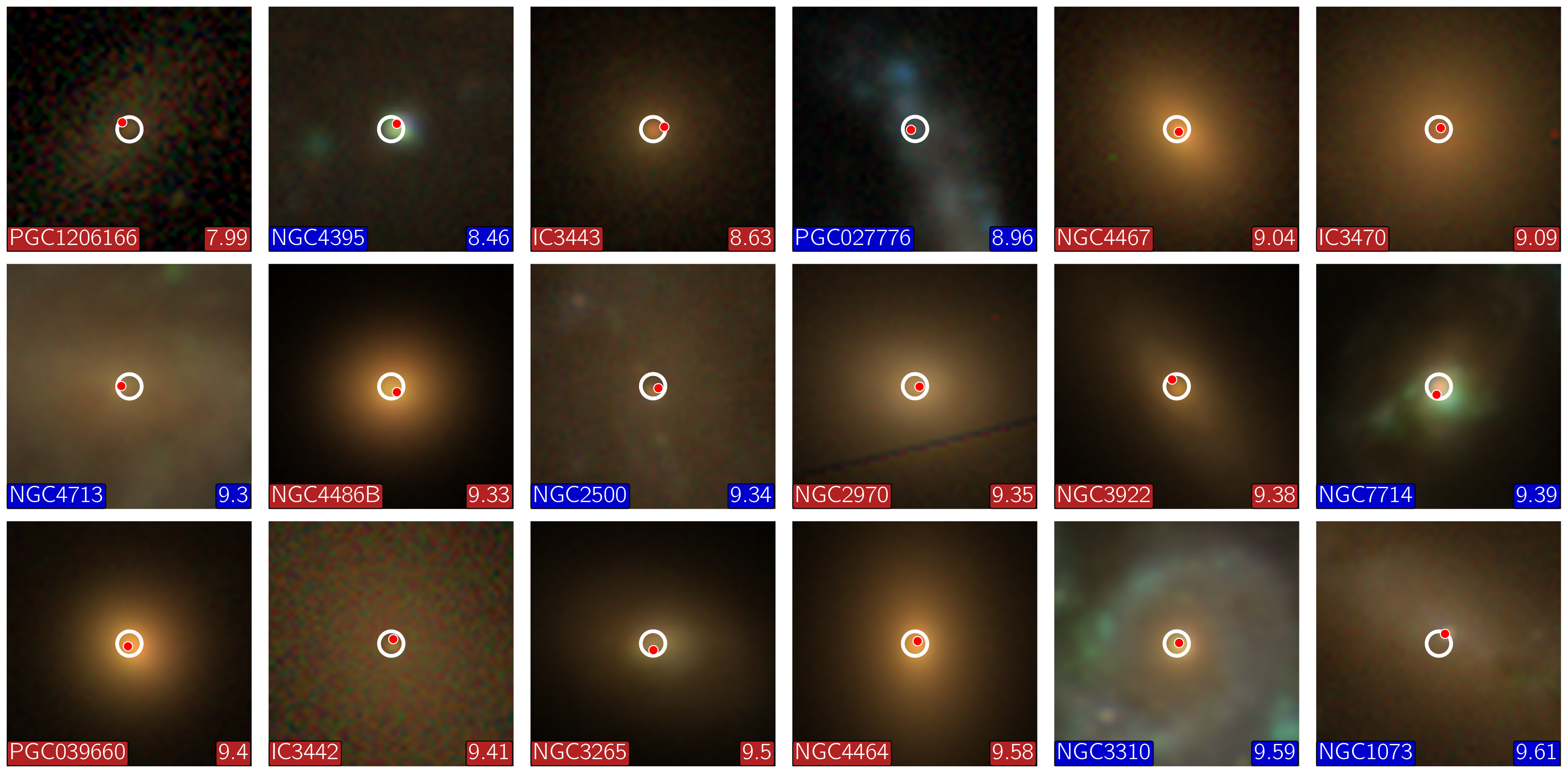}
    \noindent\rule[1ex]{15cm}{1.5pt}
    \plotone{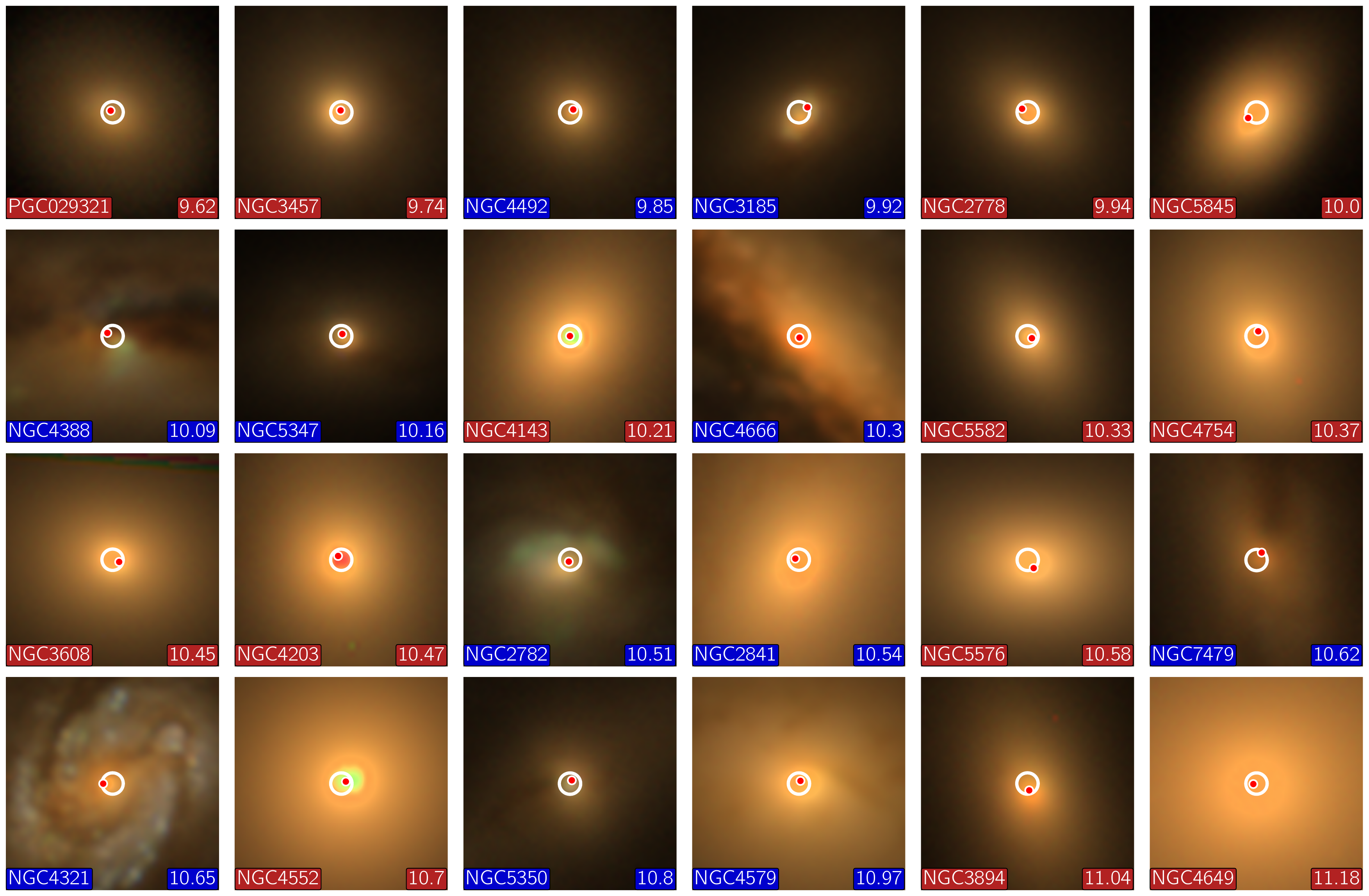}

    \caption{Images of catalog galaxies with Chandra sources within 1\arcsec\ of their centers that have log(L$_X$)$>$38.3 and have available SDSS imaging. The white circle is centered on the catalog's galaxy coordinates and shows the 1\arcsec~ radius used to match to Chandra detections, while the red dot marks the Chandra source position.  Each image has a width of 20\arcsec.  The galaxies are arranged in order of increasing log($M_{\star}/M_\odot$).  The top three rows show the 18 galaxies with the lowest log($M_{\star}/M_\odot$), and represent 18/22 lowest mass galaxies with X-ray sources.
    The bottom four rows show galaxies at higher masses sampled evenly in log($M_{\star}/M_\odot$). Image labels provide catalog {\tt objname}, {\tt logmass}, and {\tt best\_type} (red are early-types, and blue are late-types).
    \label{fig:imagesample}}
    
\end{figure*}

Of the 249 galaxies used for our active fraction analysis, 159 have available SDSS imaging. For visualizing the X-ray sources in our sample, we show 42 of these galaxies in Fig.~\ref{fig:imagesample}. In the first three rows of the figure we show the 18 lowest-mass SDSS galaxies, and a sampling of galaxies at higher masses in the bottom half.  The 18 lowest mass galaxies have log($M_{\star}/M_{\odot}$) from 7.99--9.61; there are an additional 4 galaxies in this mass range without available SDSS imaging.

\subsubsection{Local Active Fraction as a Function of Galaxy Mass and Type}

\begin{figure*}[!ht]
\gridline{\fig{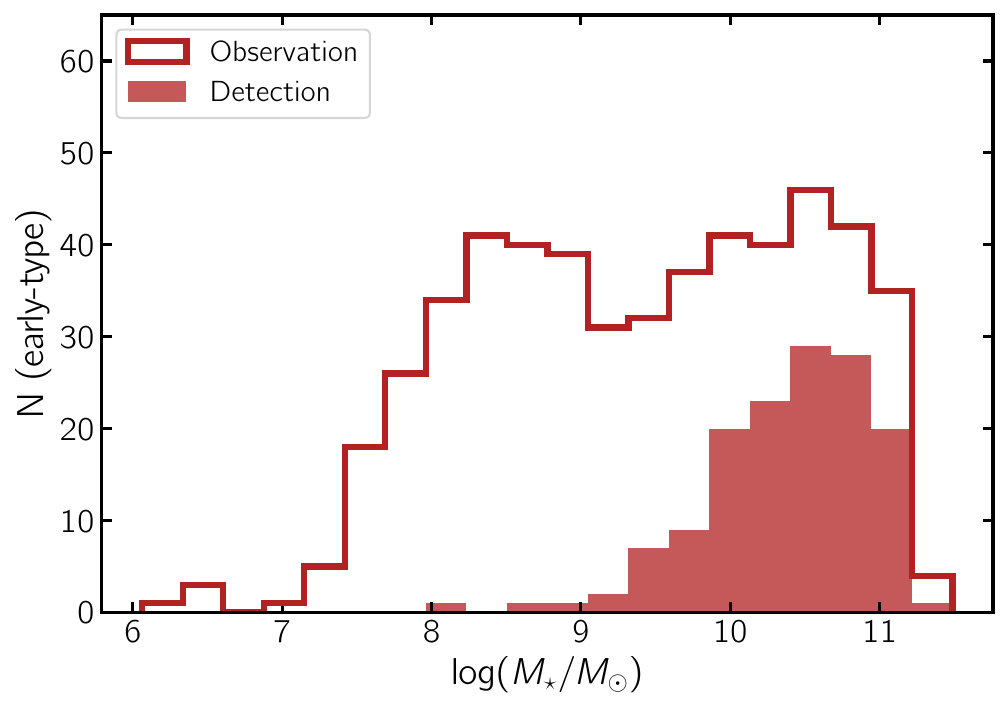}{0.45\textwidth}{(a)}
          \fig{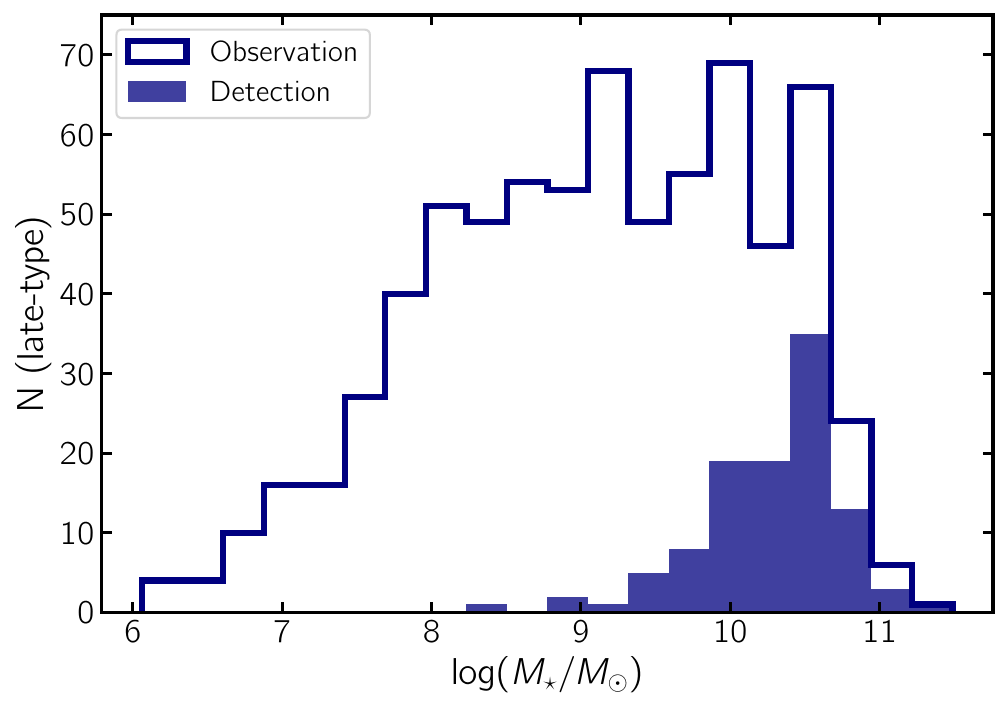}{0.45\textwidth}{(b)}}
\gridline{\fig{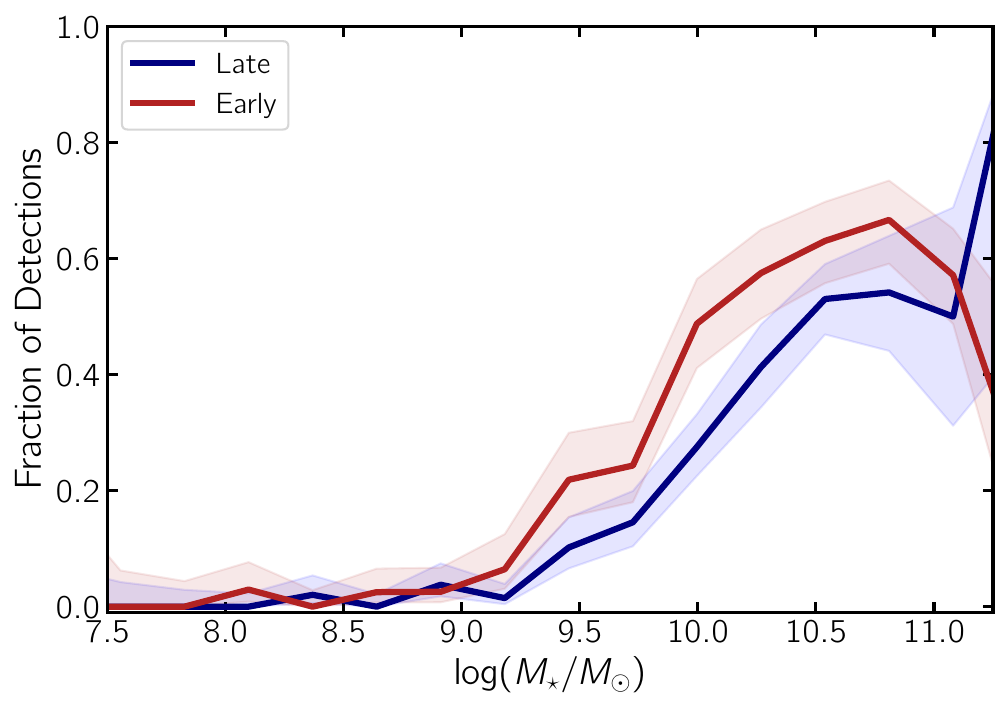}{0.45\textwidth}{(c)}
          \fig{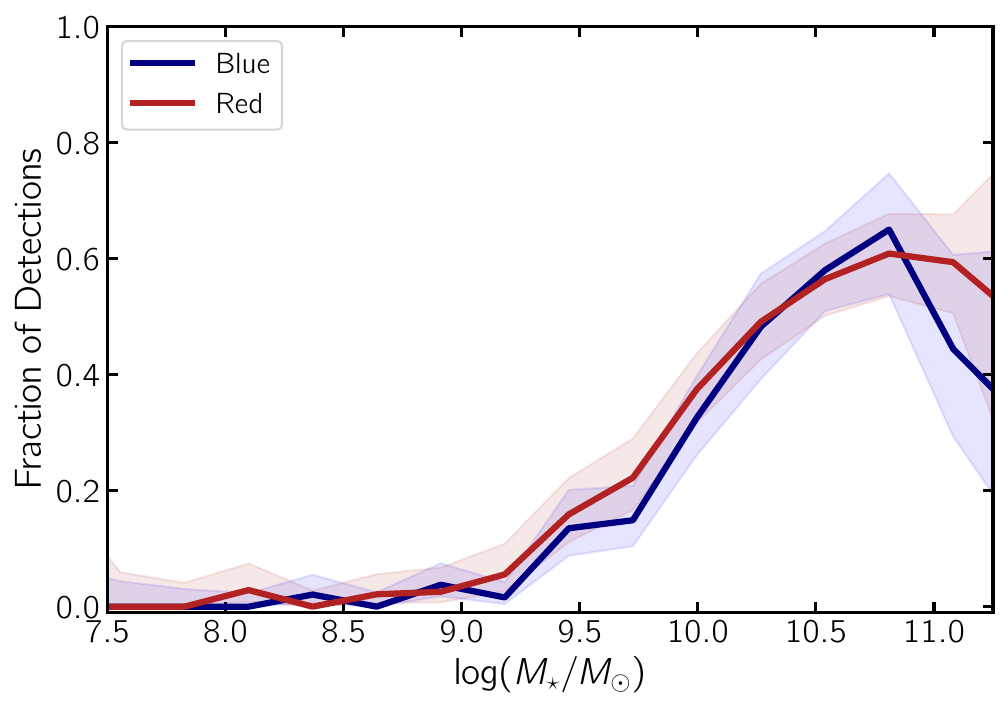}{0.45\textwidth}{(d)}}
\caption{
X-ray active fraction of catalog galaxies as a function of galaxy mass and type, using a 1\arcsec matching radius.  {\em Top row} -- the log($M_{\star}/M_{\odot}$) distribution of catalog galaxies with Chandra data. The solid line shows galaxies with Chandra observations, and the filled histogram shows galaxies with Chandra detections with log(L$_X$)$>$38.3 for {\em (a)} early-types and {\em (b)} late-types. {\em Panel c --} X-ray detection fraction as a function of log($M_{\star}/M_{\odot}$) and separated by {\tt best\_type}, a combination of morphological and color-based types  (see \S\ref{sec:type}).
{\em Panel d --} X-ray detection fraction as a function of log($M_{\star}/M_{\odot}$) and separated using {\tt color\_type}. Shaded areas in bottom figures show Agresti-Coull binomial confidence intervals.
}
\label{fig:agnfraction}
\end{figure*}

We calculate the X-ray active fraction by dividing the number of galaxies with Chandra detections by the number of galaxies with Chandra observations.  We use a 1\arcsec~matching radius, to limit the impact of X-ray binary contamination and take advantage of Chandra's excellent resolution \citep[e.g.][]{gallo10}, however, we examine the results of larger matching radii in the next subsection.
Results of this X-ray active fraction binned by mass and separated by morphological type, are shown in Fig.~\ref{fig:agnfraction}. Although some detections may be caused by nuclear X-ray binaries, this fraction serves as an upper limit of AGN occupation for our catalog.

We observe an apparent difference in detection fraction between early- and late-type galaxies, particularly within the mass range $9 \lesssim log(M_{\star}/M_{\odot}) \lesssim 10.5$. We examine the significance of this difference following the methods of \citet{hoyer21}. We calculate $p$-values under the null hypothesis that both early- and late-type galaxies have the same X-ray active fraction. We split data into 0.25 dex bins from $7.75 \leq log(M_{\star}/M_{\odot}) \leq 11.25$. We use the total X-ray active fraction in each bin as an estimator of the binomial success probability. Using this fraction, we draw $n$ samples from a binomial distribution where $n$ is the number of galaxies per bin in both the early- and late-type subsamples. We repeat this exercise $10^{6}$ times, and use the fraction of estimates that match or exceed our observed difference between early- and late-type galaxies to estimate a $p$-value for each bin. Using Fisher’s method (Fisher 1992), we combine all $p$-values into a single parameter, under the assumption that the null hypothesis is true and that the data in each mass bin are independent. The final $p$-value is 0.0004 over the entire mass range, thus suggesting that the observed enhancement of early-type galaxies is significant at a level $>$3$\sigma$.  This significance increases if, for instance, we consider just the galaxies between $9 \leq log(M_{\star}/M_{\odot}) \leq 10.5$.  Thus, it appears there is a significant enhancement of X-ray detections in early-type galaxies versus late-type galaxies. We examine whether this result is due to a higher AGN fraction in early-type galaxies below.

\subsubsection{Local Active Fraction Uncertainties and Systematics}\label{subsec:activesystematics}

\begin{figure*}
\plottwo{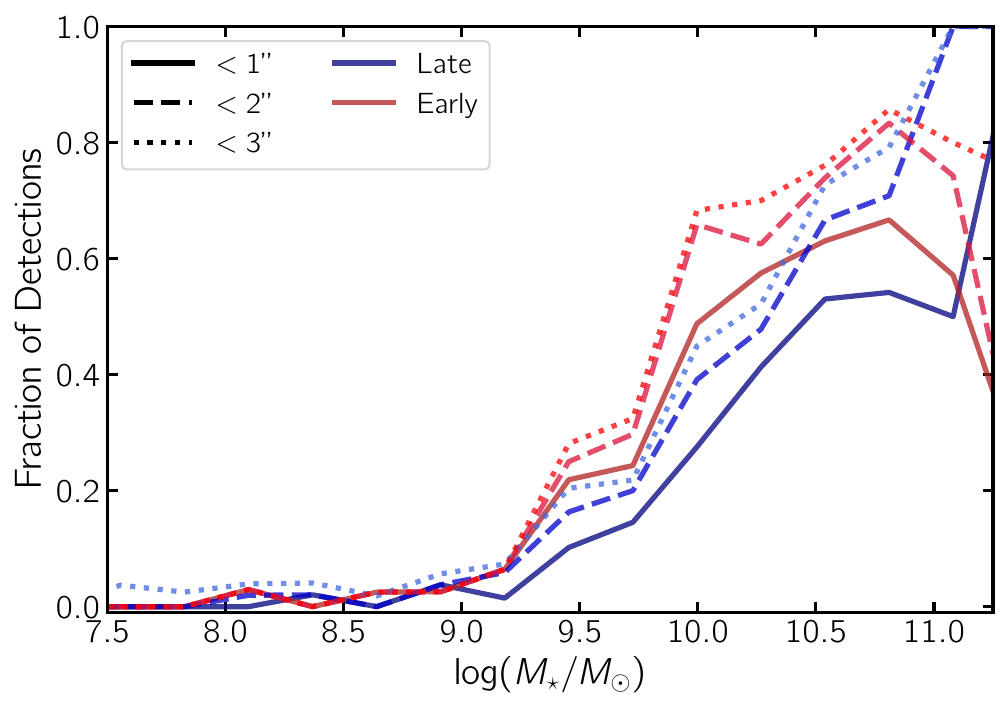}{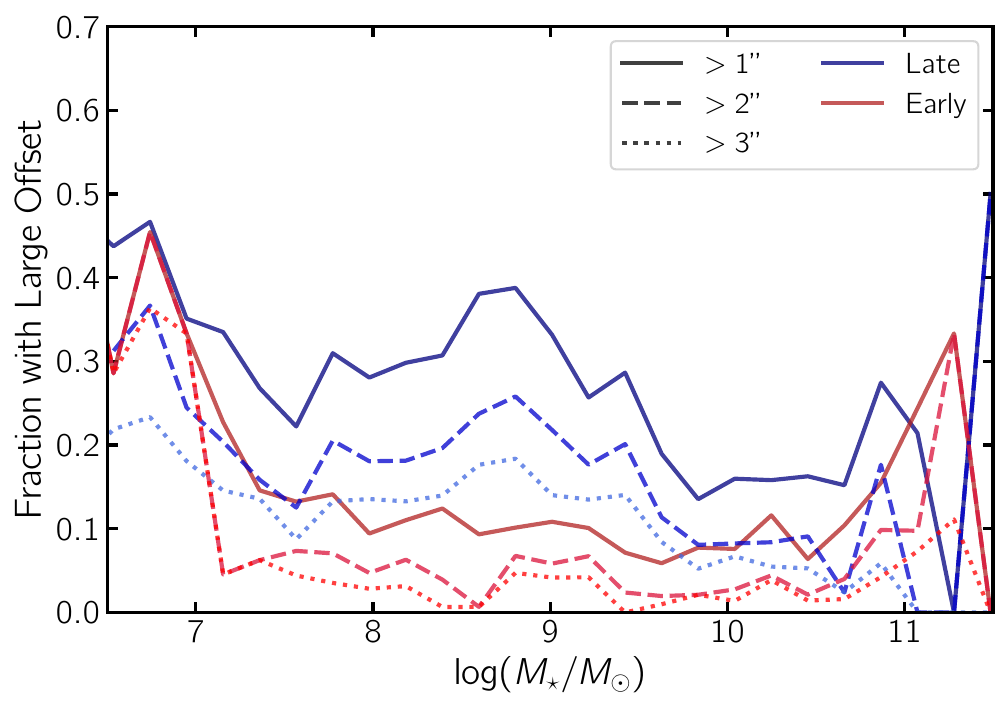}
\caption{
Position offsets and their potential effects on active fraction as a function of galaxy mass and type. {\em Left --} Comparison between the X-ray active fractions calculated using different separation radii to crossmatch with Chandra observations. 
The solid line shows active fraction using a matching radius of $1~\arcsec$, dashed lines use a radius of $2~\arcsec$, and dotted lines with a radius of $3~\arcsec$.  Red and blue lines show early- and late-type galaxies respectively. {\em Right --} Fraction of catalog galaxies with large position offsets between our catalog and NED as a function of mass and type. The solid line shows fraction with offset $>1~\arcsec$, dashed lines with an offset $>2~\arcsec$, and dotted lines with offset $>3~\arcsec$. Large offsets are seen much more commonly in late-type galaxies than in early-type galaxies, especially at lower masses.  
}
\label{fig:position_offsets}
\end{figure*}

One significant potential uncertainty is the locations of galaxy centers, and whether our catalog coordinates accurately reflect the true center of each galaxy.  To gauge the impact of this uncertainty, we conducted the same analysis using a 2\arcsec~ and 3\arcsec~ separation constraint for Chandra detection matching. As shown in the left panel of Fig.~\ref{fig:position_offsets}, larger matching radii increase the X-ray active fraction for both types overall. The difference between early- and late-type active fraction depicted in Fig.~\ref{fig:agnfraction} still exists for larger separation constraints, but the statistical significance decreases to $p=0.0019$ for 2\arcsec~ and $p=0.0104$ for 3\arcsec.

The decreasing difference in X-ray active fractions between early- and late-types with larger matching radii could be explained if the centers of late-type galaxies were less certain due to their lower surface brightness and more complicated morphology. The central surface brightness does depend on galaxy mass \& luminosity \citep[e.g.][]{ferrarese06, blanton03}, with decreasing surface brightness in lower mass galaxies below roughly the Milky Way's luminosity/mass.  Late-type galaxies also generally have lower surface brightness than early-type galaxies \cite[e.g.][]{blanton03}. In addition, many of our late-type galaxies are irregulars; these dominate our late-type galaxy number counts below log($M_{\star}/M_\odot$)$=$8.0.  Finally, distance plays a role in our ability to accurately locate the angular center of our galaxies, with more distant galaxies likely having more accurate angular coordinates.

As noted in \S\ref{subsec:combining}, we see significant discrepancies between our catalog positions (mostly from HyperLeda) and the best available positions from NED. These offsets can give us a sense of how uncertain the nuclear positions are of each galaxy. In the right panel of Fig.~\ref{fig:position_offsets}, we show the fraction of galaxies with offsets larger than 1\arcsec, 2\arcsec, and 3\arcsec~ as a function of galaxy type (early vs.~late) and stellar mass. The fraction of galaxies with significant position differences between NED and our catalog is noticeably larger for late-type galaxies than early-type galaxies in the mass range $7.5 \lesssim log(M_{\star}/M_{\odot}) \lesssim 9.5$.   For a 1\arcsec~position match, the differences are large enough ($>$20\% at some masses between early- and late-types), to explain the lower active fractions we see in late-type galaxies.  These larger positional uncertainties are likely tied to the lower surface brightness and irregular morphology amongst our late-type galaxies. 
However, at higher masses ($9.5 \lesssim log(M_{\star}/M_{\odot}) \lesssim 10.5$), the positional uncertainty differences become smaller than the active fraction differences.  
To further examine the possibility of uncertain nuclear coordinates impacting our active fraction measurements, we visually examined all 17 galaxies (16 of which are late-type) with $log(M_{\star}/M_{\odot}) \lesssim 9.5$ that had X-ray sources within 3\arcsec\ of their centers, but not within 1\arcsec.  In 11 of these 17 galaxies, the X-ray sources are clearly offset from the galaxy centers.  In another five galaxies it was ambiguous whether the X-ray source was nuclear or not, while in only one galaxy (NGC~598/M33), it was clear that the X-ray source was nuclear and our nuclear position is offset by $\sim$1\arcsec from the true center of the galaxy (this galaxy has a known nuclear star clusters that is visible in SDSS imaging).
This shows that accurate nuclear positions are an important factor to consider when examining the presence of AGN in nearby galaxies, especially in lower-mass late-type galaxies. However, it doesn't appear that errors in nuclear positions fully account for the difference we see between early and late-type galaxy active fractions.

\subsubsection{Local Active Fraction as a Function of Galaxy Mass and Color}

We also repeated the process to calculate X-ray active fractions using only color information to separate early- and late-type galaxies (using the {\tt color\_type} described in \S\ref{subsec:morph}). This is shown in the lower-right panel of Fig.~\ref{fig:agnfraction}.  When classifying type based only on color, X-ray active fractions appear similar in all mass bins. This change from our primarily morphological-based types results from a suppression of the X-ray active fraction in red galaxies relative to early-type galaxies.  This is primarily caused by galaxies classified as late-type, but with red color which lie above our color-based type separation (Fig.~\ref{fig:colormass}). Visual inspection shows that the majority of these red late-type galaxies are dusty spiral galaxies, many of which are viewed at high inclinations. This suggests that despite the potential issues with visual morphological classification (see \S\ref{subsec:morph}), that they provide valuable information on the galaxy type over and above the color-information. This is particularly true for red late-type galaxies, which are quite numerous and have accurate morphological classifications. 

The active fraction measurement we present here is the first step in a more sophisticated occupation fraction analysis (including accounting for XRB contamination) that will be presented in a follow-up paper (Gallo et al., {\em in prep}).

\section{Conclusions} \label{sec:conclusion}

In this paper, we present a robust catalog of 15424 nearby galaxies within 50 Mpc with self-consistent distance, mass, and morphological type measurements.  This catalog was constructed primarily to enable X-ray active fraction and occupation fraction measurements, but the utility of this catalog spans a much broader range of possible applications.  In particular, it provides a more complete catalog of lower mass galaxies than were included in recent similar studies \citep{she17a,bi20}.

Our catalog combines galaxies from HyperLeda, the NASA-Sloan Atlas, and the Local Volume Galaxy catalog. We also include additional photometry from Siena Galaxy Atlas and NED.
For galaxies with data from multiple sources, we compared values to the literature and include best possible measurements in the catalog. We extensively cleaned the catalog, including individual investigation into extreme outliers. Through individual investigation of SDSS imaging for NSA non-matches, we discovered that $\sim13\%$ of nearby NSA galaxies are contaminants.

We compile extinction-corrected magnitude, luminosity, and color information for 11674 galaxies (76\% of our full sample). We use the galaxy colors to estimate masses by creating self-consistent color – mass-to-light ratio relations in four bands. We also provide errors on our mass estimates that include both the scatter around our color-$M/L$ relations and distance errors. We also compare our masses to photometric masses from S4G and dynamical masses from ATLAS\textsuperscript{3D}.

Galaxy morphologies are available for 13744 galaxies.  We combine this information with galaxy masses and colors to fit a line that optimizes the separation of early- and late-type galaxies.  Through visual inspection, we find that early-type galaxies bluer than this line are frequently incorrectly classified, and we therefore reclassify those that fall more than $\sim0.13$ mags below our morphological separation line. Red late-type galaxies are typically truly late-type; many of these are edge-on dusty galaxies.

In addition we:
\begin{itemize}
    \item combine best estimates for flow-corrected redshift-based distances with redshift independent distances, with a focus on minimizing distance error, and including special treatment for galaxies in the Virgo Cluster.  
    \item use galaxies with overlapping color information to calculate and provide empirical transformations of $B-V$, $B-R$, and $g-r$ colors into Sloan $g-i$.
    \item identify galaxies in the catalog which belong to dense groups with the help of the \citet{lambert20} group catalog.
\end{itemize}

Lastly, we present a preliminary analysis of X-ray sources and find that 249 out of the 1506 galaxies with existing Chandra observations have X-ray detections with log(L$_X$)$>$38.3 and available stellar mass estimates. We present evidence that the X-ray active fractions for early- and late-type galaxies differ. In particular, we find the active fraction is higher for early-types within the mass range $9<$log($M_{\star}/M_{\odot}$)$<10.5$. We show that increased astrometric uncertainties in late-type galaxies than early-types may be partially responsible for this difference in active fractions by galaxy type. In an upcoming paper (Gallo et al., {\em in prep}) we use this catalog to conduct the first robust analysis of AGN occupation fraction down to log($M_{\star}/M_{\odot}$)=8 both overall, and separated by morphological type.

\acknowledgments

We would like to thank Michael Blanton for his useful clarifications on NSA photometry. We also thank Ian Steer for helpful correspondence and assistance concerning NED-D. A.C.S. acknowledges support from NSF grant AST-2108180.

This research has made use of the NASA/IPAC Extragalactic Database (NED),which is operated by the Jet Propulsion Laboratory, California Institute of Technology, under contract with the National Aeronautics and Space Administration.

We acknowledge the usage of the HyperLeda database (http://leda.univ-lyon1.fr).

We acknowledge the usage of the Siena Galaxy Atlas, based on the Legacy Survey. The Legacy Surveys consist of three individual and complementary projects: the Dark Energy Camera Legacy Survey (DECaLS; Proposal ID \#2014B-0404; PIs: David Schlegel and Arjun Dey), the Beijing-Arizona Sky Survey (BASS; NOAO Prop. ID \#2015A-0801; PIs: Zhou Xu and Xiaohui Fan), and the Mayall z-band Legacy Survey (MzLS; Prop. ID \#2016A-0453; PI: Arjun Dey). DECaLS, BASS and MzLS together include data obtained, respectively, at the Blanco telescope, Cerro Tololo Inter-American Observatory, NSF’s NOIRLab; the Bok telescope, Steward Observatory, University of Arizona; and the Mayall telescope, Kitt Peak National Observatory, NOIRLab. Pipeline processing and analyses of the data were supported by NOIRLab and the Lawrence Berkeley National Laboratory (LBNL). The Legacy Surveys project is honored to be permitted to conduct astronomical research on Iolkam Du’ag (Kitt Peak), a mountain with particular significance to the Tohono O’odham Nation\footnote{full acknowledgment available at \\ \href{https://www.legacysurvey.org/acknowledgment/\#siena-galaxy-atlas}{\tt https://www.legacysurvey.org/acknowledgment/\\\#siena-galaxy-atlas}}.

The Siena Galaxy Atlas was made possible by funding support from the U.S. Department of Energy, Office of Science, Office of High Energy Physics under Award Number DE-SC0020086 and from the National Science Foundation under grant AST-1616414.

\clearpage
\appendix
\section{Full Catalog}
\begin{deluxetable*}{lcccclccccc}[ht]
\label{tab:fullcatalog}
\caption{Stub of catalog, shortened to include only the most important columns}
\tabletypesize{\scriptsize}
\tablehead{
    \colhead{{\tt objname}} & 
    \colhead{{\tt ra}} & 
    \colhead{{\tt dec}} & 
    \colhead{{\tt v\_h}} & 
    \colhead{{\tt t\_type}} & 
    \colhead{{\tt best\_type}} & 
    \colhead{{\tt bestdist}} & 
    \colhead{{\tt bestdist\_error}} & 
    \colhead{{\tt logmass}} & 
    \colhead{{\tt logmass\_error}} & 
    \colhead{{\tt logmass\_src}} \\
    \colhead{} & 
    \colhead{(deg)} & 
    \colhead{(deg)} & 
    \colhead{(km/s)} & 
    \colhead{} & 
    \colhead{} & 
    \colhead{(Mpc)} & 
    \colhead{(Mpc)} & 
    \colhead{($M_{\odot}$)} & 
    \colhead{($M_{\odot}$)} & 
    \colhead{}
}
%\rotate
\startdata
NGC2788B & 135.898976 & -67.969798 & 1444.9 & 3.2 & late & 17.451 & 6.903 & 9.1114 & 0.3634 & B-V \\
NGC2793 & 139.194264 & 34.431685 & 1724.49 & 8.7 & late & 26.016 & 7.16 & 9.1861 & 0.2454 & g-i \\
NGC2798 & 139.345353 & 41.99981 & 1726.0 & 1.1 & late & 26.745 & 6.563 & 9.983 & 0.2176 & g-i \\
NGC2799 & 139.379054 & 41.994172 & 1870.86 & 8.5 & late & 30.689 & 6.964 & 9.3076 & 0.2006 & g-i \\
NGC2805 & 140.084902 & 64.102968 & 1730.35 & 6.9 & late & 27.77 & 6.506 & 9.8968 & 0.2073 & g-i \\
NGC2810 & 140.518791 & 71.843972 & 3479.7 & -4.9 & early & 51.345 & 6.415 & 10.8048 & 0.1354 & g-r \\
NGC2811 & 139.046301 & -16.312679 & 2355.6 & 1.1 & late & 28.1 & 1.29 & 10.617 & 0.1354 & B-V \\
NGC2814 & 140.297846 & 64.25236 & 1592.0 & 3.1 & late & 25.786 & 6.638 & 9.1279 & 0.2287 & g-r \\
NGC2815 & 139.082217 & -23.633299 & 2545.3 & 2.9 & late & 36.977 & 5.217 & 10.702 & 0.1354 & B-V \\
NGC2820 & 140.441039 & 64.257886 & 1575.01 & 5.3 & late & 25.507 & 6.642 & 9.7143 & 0.2315 & g-i \\
\enddata
\tablecomments{
{\tt objname} - Common galaxy name;
{\tt ra} - Right Ascension from combined sources;
{\tt dec} - Declination from combined sources;
{\tt v\_h} - Heliocentric radial velocity, see Section 3.1;
{\tt t\_type} - Numerical Hubble T-Type;
{\tt best\_type} - Combined galaxy type, see Section 5.1;
{\tt bestdist} - Our chosen best distance estimate;
{\tt bestdist\_error} - error for best distance estimate;
{\tt logmass} - Compiled best log$(M_{*}/M{\odot})$ estimate;
{\tt logmass\_error} - Error on logmass;
{\tt logmass\_src} - Color used for best mass estimate: $g-i$, $g-r$, $B-V$, $B-R$;\\
\\
\textbf{Columns found in full catalog, available from the publisher and at \href{https://github.com/davidohlson/50MGC}{\tt https://github.com/davidohlson/50MGC}}:
{\tt pgc} - PGC number (from HyperLeda);
{\tt nsa\_id} - NASA-Sloan Atlas NSAID number;
{\tt group\_id} - Galaxy is member of group, number based on Lambert et al. 2020 group catalog;
{\tt ra\_nsa} - Right Ascension provided by NSA;
{\tt dec\_nsa} - Declination provided by NSA;
{\tt ra\_ned} - Primary Right Ascension provided by NED;
{\tt dec\_ned} - Primary Declination provided by NED;
{\tt d25} - Apparent Diameter from Karachentsev and HyperLeda;
{\tt v\_cmb} - Radial velocity with respect to CMB radiation;
{\tt v\_source} - Original source for compiled velocity: HyperLeda, LVG, NSA;
{\tt hl\_obj} - True for objects in HyperLeda;
{\tt lvg\_obj} - True for objects in Karachentsev's Catalog of Local Volume Galaxies;
{\tt nsa\_obj} - True for objects in NASA-Sloan Atlas;
{\tt sga\_obj} - True for objects in Siena Galaxy Atlas;
{\tt color\_type} - Color-based Type, see Section 5.1;
{\tt a\_B\_leda} - B-band extinction from HyperLeda multiplied by 0.86 to translate to Schlafly, $A_{V}=0.769*a_{B_{leda}}$ and $A_{R}=0.629*a_{B_{leda}}$;
{\tt a\_g\_nsa} - g-band extinction from NASA-Sloan Atlas, the i-band extinction used is $=0.550*a_{g_{nsa}}$;
{\tt EBV\_irsa} - $E(B-V)$ value from IRSA dust website Schlafly values for all galaxies.  Used only for the Siena Galaxy Atlas sources, $A_{g}/EBV_{sga}=3.303$, $A_{r}/EBV_{sga}=2.285$;
{\tt Bt0\_leda} - Extinction corrected total B band magnitude from HyperLeda;
{\tt BV\_color\_leda} - $(B-V)$ color from HyperLeda;
{\tt B\_lum} - B-band Luminosity — derived for non-HyperLeda sources as described in Section 2.2;
{\tt gi\_color\_nsa} - Extinction corrected $(g-i)$ color from NASA-Sloan Atlas;
{\tt i\_lum\_nsa} - i-Band Luminosity, calculated using $M_{i,\odot}=4.53$;
{\tt gr\_color\_sga} - Extinction corrected $(g-r)$ color from Siena Galaxy Atlas;
{\tt r\_lum\_sga} - r-band Luminosity, calculated using $M_{r,\odot}=4.65$;
{\tt BR\_color\_ned} - Extinction corrected $(B-R)$ color from NED;
{\tt R\_lum\_ned} - R-band Luminosity from NED, calculated using $M_{R,/odot}=4.60$;
{\tt BMag} - Estimated absolute B-Band Magnitude for all galaxies;
{\tt gi\_color} - Estimated $g-i$ color for all galaxies with color measurements;
{\tt mag\_flag} - Flags galaxies with magnitude preference exceptions as described in Section 2.3;
{\tt cf3\_dist} - Distances from CosmicFlows3 Calculator;
{\tt cf3\_dist\_error} - Error on cf3\_dist;
{\tt zind\_dist} - Redshift independent distances;
{\tt zind\_dist\_error} - Error on zind\_dist;
{\tt zind\_indicator} - Redshift independent distance indicator listed in NEDD;
{\tt bestdist\_method} - General method used for best distances: CF3-Z, Karachentsev, NED-D, Mei, Cantiello, EVCC, HyperLeda;
{\tt bestdist\_source} - Source/Reference for bestdist, see Section 3;
{\tt dist\_ned\_flag} - Flags galaxies with NED-D best distance exceptions as described in Section 3.2;
{\tt logmass\_gi} - log$(M_{*}/M{\odot})$ from $(g-i)$;
{\tt logmass\_gr} - log$(M_{*}/M{\odot})$ from $(g-r)$;
{\tt logmass\_BV} - log$(M_{*}/M{\odot})$ from $(B-V)$;
{\tt logmass\_BR} - log$(M_{*}/M{\odot})$ from $(B-R)$;
{\tt chandra\_observation} - True if CSCView crossmatch returned a limiting sensitivity at matching position as described in Section 6;
{\tt chandra\_detection} - True if CSCView crossmatch returned flux information within a 1\arcsec\ matching radius as described in Section 6;
{\tt log\_lx} - X-ray luminosity log(L$_X$/erg/sec) calculated from the CSC 0.5-7 keV flux, using a 1\arcsec\ radius;
{\tt chandra\_detection\_3arcsec} - True if CSCView crossmatch returned flux information within a 3\arcsec\ matching radius as described in Section 6;
{\tt log\_lx\_3arcsec} - X-ray luminosity log(L$_X$/erg/sec) calculated from the CSC 0.5-7 keV flux, using a 3\arcsec match radius.
}
\end{deluxetable*}

\clearpage
\bibliography{bibliography}{}
\bibliographystyle{aasjournal}

\end{document}